# PULSATING WHITE DWARF STARS AND PRECISION ASTEROSEISMOLOGY

D.E. Winget[1,2] and S.O. Kepler[2]

[1]*Department of Astronomy and McDonald Observatory, University of Texas, Austin, Texas 78712; email: dew@astro.as.utexas.edu*

[2]*Instituto de Física, Universidade Federal do Rio Grande do Sul, 91501-970 Porto Alegre, RS - Brasil; email: kepler@if.ufrgs.br*

■ **Abstract** Galactic history is written in the white dwarf stars. Their surface properties hint at interiors composed of matter under extreme conditions. In the forty years since their discovery, pulsating white dwarf stars have moved from side-show curiosities to center stage as important tools for unraveling the deep mysteries of the Universe. Innovative observational techniques and theoretical modeling tools have breathed life into precision asteroseismology. We are just learning to use this powerful tool, confronting theoretical models with observed frequencies and their time rate-of-change. With this tool, we calibrate white dwarf cosmochronology; we explore equations of state; we measure stellar masses, rotation rates, and nuclear reaction rates; we explore the physics of interior crystallization; we study the structure of the progenitors of Type Ia supernovae, and we test models of dark matter. The white dwarf pulsations are at once the heartbeat of galactic history and a window into unexplored and exotic physics.

## 1. A TASTE OF HISTORY

The history of science contains blueprints for the future. We see common threads running through much of the stories of past progress and discovery. Looking back at the history of scientific achievement, we see this theme repeated many times: meaningful progress requires that theory and experiment advance together. They must coordinate like two legs of a child learning to walk. If one leg gets too far ahead, the child falls. To move forward again, the child is forced to gather both legs together to stand. Examples when one leg gets farther in front are numerous in all fields of science. An early astronomical example is the Greek idea that the Earth might orbit the Sun. Searching for and not finding parallax, they abandoned the idea----the observational techniques of the time were not adequate. More recently, the cosmic background radiation might have been discovered in the 1940s through the observations of interstellar lines, were it not for lack of a theoretical context.

Our analogy works in another way: It doesn't matter too much which leg takes the first step as long as they stay close enough together. Kepler built his laws on Tycho's careful observations; Newton "stood on the shoulders of giants;" Einstein's general relativistic description of gravity required Eddington and a total solar eclipse to be taken seriously.

The attentive reader will see many parallels throughout this review between the quantum theory of the atom and white dwarf stars. The structure of both is determined through the delicate balance of a strong central force against the uncertainty and exclusion principles. Not surprisingly, their stories are intertwined; the white dwarf stars have a long history as a cosmic laboratory for quantum physics and general relativity.

We are not historians; the science we review is of a more humble nature than these lofty examples. We offer the history of the field---from the inevitably biased viewpoint of your guides---as an introduction and a starting point for a blueprint of the future. The study of pulsating stars is more than 300 years old; the pulsating white dwarf stars are relative newcomers to this venerable field. The first step was taken by observers.

Our story begins with a man, a star, and a puzzle. The man was a careful observer named Arlo Landolt. In 1964, he was making observations as part of a major effort to establish a web of photometric standard stars on the sky. In his observations of one particular star, he found small residuals that varied irregularly on a rapid timescale. A less careful observer might have written the residuals off as errors. A less conscientious one might have been tempted to toss the one star from his large sample, forgetting about its small variations. Fortunately, Landolt did neither.

He reobserved the star, now known as HL Tau 76, photometrically and spectroscopically over the next three years. In the process he established two properties: it varied with at least one period near 12.5 min, and it was consistent with a single white dwarf star---not a binary ([Landolt 1968](#)). Taken together, these results were quite unexpected and difficult to reconcile. The timescale for the variations was within the range observed in known blue variables containing white dwarf stars, but all those were binaries with variations attributable to rotation or orbital periods. It was well established that the timescale for radial pulsations of an isolated white dwarf star was of order a few seconds and possibly less ([Sauvenier-Goffin 1949](#), [Schatzman 1961](#), [Ostriker & Axel 1969](#)). [Eddington (1926)](#) pointed out that the radial pulsation timescale is set by the sound travel time across the star, or equivalently (using the virial theorem), by the free-fall timescale,

$$\tau_{\text{free-fall}} \sim \tau_{\text{dynamical}} \sim \frac{1}{(G\bar{\rho})^{\frac{1}{2}}} \sim \text{seconds, for a white dwarf}.$$

This is clearly more than two orders of magnitude too short to explain the observed variations. Here was the rather formidable puzzle. Landolt went to meetings to tell others about his puzzle, hoping for an answer. The solution created the field we review. To understand the solution, we need a brief introduction to asteroseismology and nonradial pulsations. Before this, we examine why we should care about a solution at all. We need to understand why white dwarf stars are important and interesting.

## 2. MOTIVATIONS AND CONTEXT

White dwarf stars are historically well established as cosmic laboratories for testing physics under extreme conditions. We briefly examine their role in this context in the past, present, and immediate future. They are the only stable evolutionary state for the overwhelming majority ($\simeq 98\%$) of all stars; they present us with a very important boundary condition for prior stellar evolution. They have become important in several cosmological contexts---as candidates for dark matter, as microlensing objects, and as the progenitors of supernovae of Type Ia (SNIa). SNIa apparently exhibit significant peak-magnitude variations with redshift. If not due to differences in the internal properties of SNIa progenitor white dwarf---possible because of differences in metallicities for the Pop. III and Pop. II stars and in the resulting white dwarf stars (Kalirai et al. 2008) ---these magnitude variations imply SNIa could be used to measure acceleration in the expansion of the universe. Finally, white dwarf stars have been recognized for more than forty years to be accurate clocks, owing to the simplicity of their evolution. Therefore they have become important in determining the cosmic chronology (Winget et al. 1987, Hansen & Liebert 2003, Hansen et al. 2004). We examine all of these aspects. The hot subdwarf stars are interesting in their own right, but beyond the scope of our review. Please see Fontaine, Brassard & Charpinet (2003) and Fontaine et al. (2006) for excellent reviews of compact pulsators that include the pulsating sdB stars.

### 2.1. Physics and the White Dwarf Stars

Large-scale experiments in physics and astronomy must sail the stormy seas of politics on a course plotted by legislators. In this environment, astronomy, with its broad popular appeal, has often fared better than physics. This has enabled some noble projects like the *Hubble Space Telescope* to navigate these troubled waters successfully, while others, such as the Super Conducting Super Collider, survived many of the hardships only to be dashed upon the rocks of budget cuts at the last possible moment.

Fortunately, nature conducts large-scale physics experiments in space, and we have only to pay the cost of observing them. White dwarf stars are an example. Here we can explore the physics of matter under a wide range of extreme conditions, well outside the domain accessible in even the best-funded terrestrial laboratory. An illustration of the large range of temperatures and pressures inside white dwarf stars appears in Figure 1.

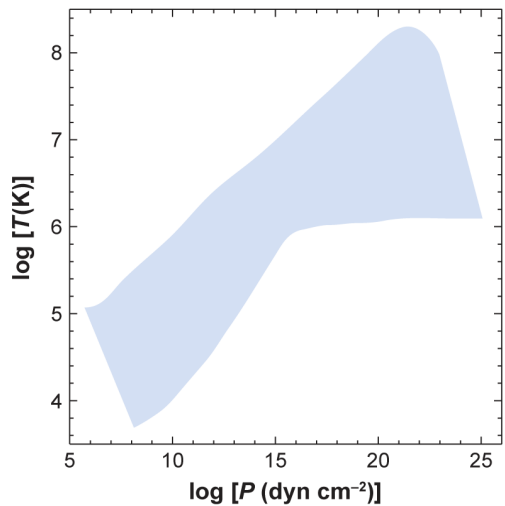

**Figure 1** The region of the Pressure-Temperature plane spanned by the white dwarf star models. Pressure is in units of dynes per square centimeter. The extreme boundary is approximately marked at the upper left by a $0.6 M_\odot$ model appropriate to the hot, pre-white dwarf stars. The lower-right boundary is for a $1.2 M_\odot$ model representing the coolest observed white dwarf stars.

Understanding the existence of white dwarf stars provided an early test of the quantum theory of matter (see Van Horn 1977 for an elegant discussion). Solving the mystery of the source of internal pressure in an object with the mass of the Sun and the size of the Earth gave us a powerful demonstration of the combined action of the Pauli exclusion principle and the Heisenberg uncertainty principle. Fowler 1926, and Chandrasekhar (1935, and references therein, showed that the speed of light limits the pressure support from degenerate electrons. Therefore white dwarf stars have a maximum possible mass. Later, we would see that white dwarf stars could provide a test not only of special relativity and quantum mechanics but also of general relativity. Strong gravity implies a detectable gravitational redshift. This was first reliably measured by Greenstein & Trimble (1967) and has become an important way to measure the masses of white dwarf stars (Reid 1996, and references therein).

A glance at Figure 1 makes evident the range of physical processes dominating the structure and evolution of the white dwarf stars. A $0.6 M_\odot$ model sequence representing the hot, pre-white dwarf stars defines the upper boundary of the region

(from Kawaler 1986), and a $1.2\,M_\odot$ model sequence representing the massive coolest white dwarf stars defines the bottom boundary (from Montgomery 1998).

At the extreme temperatures expected in hot, pre-white dwarf stars, the energy loss by photons is completely swamped by neutrino losses. These are a factor of 10 larger. The resultant temperature inversion is evident in the hot model. The temperature maximum is not at the central density maximum, because the neutrinos refrigerate the deep central regions. This unusual temperature gradient implies a negative photon luminosity (inward photon flux) in central regions of these models.

Many types of neutrinos are expected in theoretical models of these stars, but plasmon neutrinos dominate the energy loss in most of the regions where neutrinos are important (see Itoh et al. 1996 and Kantor & Gusakov 2007 for theoretical rates).

[Plasmon neutrinos are the decay results from a photon coupled to the plasma.]

This is the environment known where we have an opportunity to study the production of plasmon neutrinos. These rates drive the evolution of the star, so a measurement of evolutionary timescales in these stars provides a direct measure of plasmon neutrino rates, thereby testing the theory of electro-weak interactions in a new domain.

As the models evolve beyond the hot, pre-white dwarf stage, photon cooling dominates and gravitational contraction is dramatically reduced as the interior equation of state hardens into that of a strongly degenerate electron gas. Mechanical and thermal properties separate (see Van Horn 1971 for a nice discussion) at these stages and below. The degenerate electrons provide the dominant pressure, while the thermal motions of the ions make a negligible and ever-dwindling contribution to the mechanical support. The roles of electrons and ions are reversed in their contribution to the energy. The heat capacity of the degenerate electrons is negligible (Mestel 1965), and the only significant source of energy is the reservoir of thermal energy in the nearly classical ideal gas of the ions. Energy transport in the interior is dominated first by neutrinos and then by conduction. In the outer layers it is dominated by radiation, whereas in some cooler temperature regions it is dominated by convection associated with the partial ionization of the most abundant element at the surface.

In the central regions where neutrinos once reigned supreme, the Coulomb interactions become increasingly important. First, the ratio of Coulomb energy to thermal ion energy exceeds one and the material becomes a Coulomb liquid. Later, as temperatures drop further, this ratio climbs to about 180, and we expect a change in the state of matter as the ions crystallize into a lattice. Nearly forty years separate the theoretical prediction of [Salpeter (1961)](), and independently [Kirzhnits (1960)]() and [Abrikosov (1961)](), that the ions should crystallize from the first empirical evidence that this effect occurs ([Winget et al. 1997]()).

Subsequent work has shown that the transition to a solid state may leave a signature not just in the structure of the white dwarf star, but also in their evolution. [Van Horn (1968)]() showed that a first-order phase transition should release latent heat, thereby providing an additional source of energy and lengthening the white dwarf cooling times. A further effect was suggested by [Stevenson (1977](), [1980]()). He investigated the possibility that carbon and oxygen interiors might not crystallize as an alloy but might undergo phase separation upon crystallization. Early work by [Barrat, Hansen & Mochkovitch (1988)](), [García-Berro et al. (1988)](), [Mochkovitch (1983)]() indicated that the phase separation---and trace element separation as well---may provide significant additional energy sources through the redistribution of the ions of differing chemical potential energy.

[Mestel & Ruderman (1967)]() pointed out that after crystallization spreads from the center toward the surface layers, the heat capacity in the crystallized regions will no longer be due to thermal particle motions in a gas but to quantized lattice vibrations in the form of phonons. When the interior temperature drops below the point where the kT energy is less than the lowest phonon energy, the Debye temperature of the solid, higher energy levels become increasingly inaccessible and the heat capacity drops. [Ostriker & Axel (1969)]() showed that the change in heat capacity associated with the solid phase should result in a much more rapid evolution after crystallization, ultimately ending in a Debye-cooling phase where the interior heat capacity drops in proportion to $T^3$.

Theoretical models indicate that further cooling has dramatic effects on the surface layers as well. As the surface temperatures drop, gravitational settling dramatically influences the distribution of chemical elements in a white dwarf star, and the chemical stratification produced by prior evolution. The heavier elements tend to diffuse toward the center. When the surface layers become partly ionized, radiative energy transport gives way to turbulent transport in the comparatively high pressures of the white dwarf envelope. As the star cools, the turbulent energy transport moves inward to ever-higher pressures. The cooling causes the degeneracy boundary in the envelope to move toward the surface. Eventually the base of the partial-ionization zone and the degeneracy boundary collide and the turbulence is halted by the efficient energy transport associated with the degenerate electrons. Depending on the relative thickness of the various layers, as the turbulent convection reaches its deepest extent, some of the effects of gravitational settling and chemical diffusion may be reversed. Heavier elements may come to the surface if the outermost layer is thin enough. This latter effect may alter the observed surface abundances as a function of the star's effective temperature.

Even the outermost layers become exotic by terrestrial standards as the temperatures drop further to the range of the coolest observed white dwarf stars. We begin to observe molecular H. Even a form of molecular He is suspected in the high-pressure atmospheres of cool He-white dwarf stars (see Bergeron, Ruiz & Leggett 1997 and references therein). Also, the star can become bluer as it cools due to collision-induced absorption (CIA), an effect clearly seen in globular cluster white dwarf stars (Richer et al. 2006).

Looking again at Figure 1, we realize that we must understand the entire shaded region if we are to model white dwarf stars. Phrased more positively, Figure 1 dramatizes what white dwarf stars can teach us about the behavior of matter under extreme conditions.

## 2.2. White Dwarf Stars as a Boundary Condition for Stellar Evolution

Observations of star clusters give us our best estimate for the upper mass limit for the main sequence stars that become white dwarf stars. Depending on the theoretical models used, the values range from 5 to $9 M_\odot$ (e.g., Ferrario et al. 2005, Williams 2007). However, the theoretical models of Iben, Ritossa & García-Berro (1997) predicted O/Ne/Mg-core white dwarf stars descending from stars with initial masses up to $10.5 M_\odot$ on the main sequence. This implies that more than 98% of all stars will eventually become white dwarf stars. From a more personal perspective, the Sun will become one. The connection with prior evolution is summarized in a remark we often hear from Brian Warner, "Inside every red giant is a white dwarf waiting to get out."

An interesting consequence of the relatively large upper limit for the initial mass is that all stars more massive than $2 M_\odot$ ultimately recycle into the interstellar medium more mass than they remove in the form of the remnant white dwarf. Although most of the recycled material is unlikely to have undergone significant nucleosynthesis, the abundances of the relatively fragile species may be dramatically affected by recycling.

The extreme difference between the initial main-sequence masses and the final white dwarf masses underscores the large role that mass-loss must play in post-main-sequence, pre-white dwarf evolution. The qualitative stages by which the star moves off the main sequence we believe to be well understood. Evolution proceeds from core H, then He, burning followed by H- and He-burning shells surrounding a C/O core. The relative masses of the remnant H-layer and He-layer, and the relative distribution of C and O in the core and just how the mass is lost are still only very poorly understood. Measurements of the H-layer and He-layer masses and the distribution of C/O in the core are therefore of great interest as boundary conditions for these complex evolutionary stages.

White dwarf stars come in a variety of spectral types, according to the dominant observable surface element. The lion's share are H-atmosphere white dwarf stars (DAs) comprising around 84% of all white dwarf stars, whereas the He-dominated DOs and DBs total most of the remaining 16% (Eisenstein et al. 2006).

**2.3. White Dwarf Stars and Cosmology**

White dwarf stars are candidate progenitors of SNIa. An understanding of their structure and dynamical properties may improve our understanding of the resultant supernova events. Effects such as crystallization, phase separation, isotopic fractionation, dynamical and pulsational modes of energy transport, and the dependence of mass loss on metallicity must be understood. They may affect our interpretation of variations in supernova brightness with redshift.

White dwarf stars in clusters have also been used to derive cluster distances, thereby affecting age estimates based on main-sequence turn-off or post-main-sequence evolution. The idea is to use the sequence of white dwarf stars observed in the cluster plus the calibration provided by a sequence of local white dwarfs to derive the cluster distance. This has been done for example for NGC 6752 (Renzini et al. 1996), M4 (Hansen et al. 2004), and Ω Cen (Calamida et al. 2008).

Accurate masses and luminosities are required for all of these applications. Asteroseismology can provide them. Hence, finding pulsators in clusters deserves a high priority.

## 2.4. White Dwarf Stars as Galactic Chronometers

Four reasons combine to make the white dwarf stars excellent chronometers for measuring the age of the various stellar populations of the Galaxy. The first reason is that white dwarfs are representative of the general population of stars; as we have seen, most stars are or will become white dwarf stars. Second, they are a homogeneous class of objects with a very narrow distribution of total stellar masses centered between $0.15 \leq M_{wd}/M_\odot \leq 1.36$ and a mean mass of $\langle M \rangle_{DA} \simeq 0.593 \pm 0.016 M_\odot$ for the 84% of all white dwarf stars that are DAs (Kepler et al. 2007). Further, on theoretical grounds (see Iben 1991 for an excellent summary), we expect that all of the objects with masses in the range 0.45--1.1 $M_\odot$ to have C/O cores with thin overlying He-layers and at most an additional surface H layer. Third, high surface gravity ($\log g \sim 8$), typically slow rotation rates and low magnetic fields, and nuclear energy and gravitational energy generation rates near zero support the idea that these objects are physically simple. Fourth, the unavailability of significant energy sources other than the residual thermal energy of the ions allow us to conclude that the evolution of white dwarf stars is primarily a simple cooling problem. Then there is a straightforward relation between the age of a white dwarf and its luminosity. These factors combine to imply that the white dwarf stars provide an archaeological record of the history of star formation in the Galaxy.

**2.4.1. WHITE DWARF EVOLUTION AS COOLING** Many years ago, Mestel (1952) demonstrated the simplicity of the evolution of a star supported by electron degeneracy pressure but without significant nuclear energy sources. He constructed an analytical model for white dwarf evolution by treating the structure as if it consisted of only two layers. The inner layer contains most of the mass of the star and is assumed to be isothermal because of strong $e^-$-degeneracy. For the same reason, the electrons do not contribute significantly to the heat capacity; it comes almost entirely from the ions, which are assumed to behave as a classical ideal gas. The thin, nondegenerate outer layer forms an insulating blanket and controls the rate at which the energy from the ion reservoir is leaked out into space. The specific rate is controlled by the radiative opacity at the boundary between these two layers and is assumed to obey Kramer's law for the opacity, $\kappa = \kappa_o \rho T^{-3.5}$, where $\rho$ is density and $T$ is temperature. The Mestel theory has been nicely summarized by Van Horn (1971). As expected, it gives a simple form for the age-luminosity relation,

$$\log(\tau_{cool}) \approx \text{Const.} - \frac{5}{7}\log(L/L_\odot).$$

Here the age is the cooling time, $\tau_{cool}$ and $L$ is the luminosity. This analytical description of white dwarf cooling is a good picture to have in your head; as shown by Iben & Tutukov (1984), it gives surprisingly good agreement with the predictions of detailed numerical models. It is often useful to approximate the key physical effects not included in the Mestel theory as perturbations.

As discussed by Lamb & van Horn (1975), we can expect five physical effects to produce deviations from the predictions of classical Mestel theory. These are neutrino energy loss, possible residual nuclear burning in the H-layer (dismissed by Lamb & van Horn 1975 and previous researchers, but see Iben & Tutukov 1984 and Iben & MacDonald 1986), gravitational contraction, surface convection, and crystallization. In the latter, we include the closely related phase separation as well as the subsequent Debye cooling (where the ion specific heat at constant volume, $C_v^{ion}$, is proportional to $T^3$).

Each of the above physical effects has specific consequences for the distribution of normal modes —this distribution can be studied asteroseismologically.

**2.4.2. The Age and History of the Galactic Disk from White Dwarf Stars**

The idea of using white dwarf stars as chronometers has a long history beginning with Mestel (1952) and Schwarzschild (1958). They made clear that if we could find the coolest white dwarf stars, we would have found the oldest white dwarf stars (allowing for mass effects), and thereby a constraint on the ages of the oldest stars. This idea was further developed by Schmidt (1959) in a galactic context and later by Dantona & Mazzitelli (1978).

Observational results (Liebert 1979; Liebert, Dahn & Monet 1988, 1989) and Winget et al. (1987) further calculations of theoretical white dwarf luminosity functions, incorporating a distribution of masses and an estimate of the average main-sequence evolutionary time, convinced them that the turndown Liebert had found was indeed a measure of the age of the galactic disk. Winget et al. (1987) published this method of cosmochronometry along with a preliminary result of the age of the Galactic disk as $\tau_{disk} \approx 9.3 \pm 2.0$ Gyr. This work also demonstrated the possibility of using the shape of the luminosity function prior to the turndown to constrain constitutive physics in the models and the star formation history of our Galaxy.

There was still a problem in this method of obtaining an age for the Galactic disk, in that there is a broad dispersion in ages for the coolest white dwarf stars depending on which group carried out the evolutionary calculations. Winget & Van Horn (1987) demonstrated that this variation was due to very different treatments of the constitutive physics (for example inclusion of crystallization, treatment of convection) or different choices of model parameters (stellar mass, or choice of surface layer masses). They demonstrated that, when these effects are taken into account, there is remarkable agreement between the results of various investigators. Some still believed, however, that the observed downturn might be the result of short cooling times and subsequent Debye cooling, as indicated by the comments of D'Antona following the paper by Winget & Van Horn (1987, p. 376).

Further theoretical calculations employing more sophisticated treatments were carried out by Noh & Scalo (1990) and Weidemann & Yuan (1989), and culminated in the work of Wood (1990, 1992, 1995). By this time, there was a consensus that the white dwarf luminosity function gave one of the most accurate measures for the age of the Galactic disk (see Peebles 1993 and references therein).

With all this effort and enormous improvements in the treatment of the evolutionary models and the constitutive physics, it is important to see how far we have come since the beginning of this field more than 40 years ago. It is amusing to compare the current results with those published for the age of the lowest luminosity white dwarfs by Schwarzschild. Leggett, Ruiz & Begeron (1998) used the Liebert data sample supplemented with new optical and infrared data and derived temperatures from the dramatically improved cool model atmospheres of Bergeron, Saumon & Wesemael (1995). Using Wood's theoretical models they obtained

$$\tau_{disk}(1988) = 8 \pm 1.5 \text{ Gyr},$$

whereas Schwarzschild, armed with the Mestel theory, obtained

$$\tau_{disk}(1958) = 8 \times 10^9 \text{ yr}.$$

With tongue in cheek, we note that in 40 years we have gained both error bars and a new expression for units. In a more serious vein, we must point out that with the same Leggett, Ruiz & Bergeron (1998) data, and the phase-separation models of Montgomery (1998) or Salaris et al. (1997), the value for the age becomes 9 Gyr. Also, starting with Hansen et al. (2002), white dwarf cooling sequences for the closest globular cluster have become available, and the ages of halo white dwarf stars indicate an age for the halo of our galaxy,

$$\tau_{halo} = 12 \pm 1.5 \text{ Gyr}.$$

Much more detail can be found in [Hansen & Liebert (2003)](#), and we recommend [Fontaine, Brassard & Bergeron (2001)](#) for an authoritative review, and the recent papers on the effects of Ne deposition by e.g. García-Berro et al. (2008).

Clearly there is much more to the story but this remarkable consistency underscores why white dwarf stars are such useful chronometers: They are simple!

## 3. A GENTLE GUIDE TO ASTEROSEISMOLOGY AND NONRADIAL PULSATIONS

### 3.1. Why Asteroseismology?

The curse of astronomy is that we cannot perform experiments. We can only observe what nature chooses to show us. When we observe stars, we see only the light from the outermost layer, the photosphere. Most of the interesting physics is inside. Asteroseismology gives us a way to look inside.

### 3.2. How Do We Do Asteroseismology?

We use the time-honored approach of physicists for centuries: We study a system by examining its normal modes. This has been a well-established tool for nearly a century in the study of the Earth's interior, hence the term seismology. Many of our ideas and approaches have been borrowed wholesale from this field. Perhaps a more familiar analogy is with atomic structure. We see from the mathematics why the problems are so analogous. Both are simple spherical-potential problems. The pulsation modes in the star are described in the same way as the energy levels of the atom.

### 3.3. The Physics of the Frequencies

In order to understand the oscillations we observe in white dwarf stars, we need to consider the general class of nonradial spheroidal oscillations. We derive the equations we actually solve by perturbing the fluid equations and keeping only the terms of lowest order. Nonlinear effects are clearly important and of increasing interest, but they are well beyond the scope of our present discussion. We assume a static, spherical equilibrium structure that is given by a theoretical evolutionary model. Because the surface gravity, $g$, is high, $\log g \sim 8$ in units of cm/s$^2$, and because rotation rates are typically on the order of days, the spherical approximation is quite good. So we can expand our solutions for the perturbative fluid displacements in terms of spherical harmonics, $Y_{\ell m}$. Further, we seek periodic solutions of the form,

$$\xi(\mathbf{r},t) = \xi(\mathbf{r})e^{i\sigma t},$$

where $\sigma$ is the frequency of the pulsation. The most relevant parameters that describe the equilibrium model are the Brunt-Väisälä frequency, N, proposed independently by Vilho (Yrjö) Väisälä in 1925 and David Brunt, in 1927, which in the absence of chemical potential gradients is given by

$$N^2 \equiv -Ag = -g\left(\frac{d\ell n\rho}{dr} - \frac{1}{\Gamma_1}\frac{d\ell nP}{dr}\right),$$

which is just the difference between the actual and the adiabatic density gradients, where $\Gamma_1$ is the adiabatic gas coefficient – the logarithmic adiabatic derivative of the pressure, P, with density, $\rho$. The Lamb, or acoustic, frequency, $S_\ell$, is given by

$$S_\ell^2 \equiv \frac{\ell(\ell+1)}{r^2}\frac{\Gamma_1 P}{\rho} = \frac{\ell(\ell+1)}{r^2}v_s^2,$$

where $v_s$ is the local sound speed (Horace Lamb 1910).

Local analysis can give us a great deal of physical insight into the oscillations and how they sample the star (see, for example, Unno et al. 1989). If we assume a radial dependence of the displacements of the form, $e^{ik_r r}$, and wavelengths short compared to the relevant scale heights for the physical quantities, we arrive at a local dispersion relation (LDR) of the form,

$$k_r^2 = \frac{k_h^2}{\sigma^2 S_\ell^2}(\sigma^2 - N^2)(\sigma^2 - S_\ell^2)$$

Here we have defined a horizontal wave number,

$$k_h^2 \equiv \frac{\ell(\ell+1)}{r^2}\frac{S_\ell^2}{v_s^2},$$

such that the total wave number is

$$k^2 \equiv k_h^2 + k_r^2.$$

The LDR allows us to see how the two characteristic frequencies together determine the nonradial pulsation properties of the star. In order for a given mode to be locally propagating, $k_r^2$ must be positive. From the above expression, we see that this occurs only when the oscillation frequency is greater than or less than both the Brunt-Väisälä frequency, $N$, and the Lamb frequency, $S_\ell$.

Taking the limits of large and small frequencies, the LDR yields two physically distinct kinds of solutions that represent the two principle classes of nonradial spheroidal modes:

1. Where $\sigma \gg N^2, S_\ell^2$:

$$\sigma_p^2 \approx \frac{k^2}{k_h^2} S_\ell^2 = (k_r^2 + k_t^2)v_s^2.$$

2. Where $\sigma^2 \ll N^2, S_\ell^2$:

$$\sigma_g^2 \approx \frac{k_h^2}{k_r^2 + k_h^2} N^2.$$

The first class of solutions represents the p-modes, so called because pressure is the principal restoring force. Radial displacements are dominant. For white dwarf stars these have timescales of seconds, too short to be the periods of the observed oscillations. An additional complication is the energy needed for the large radial displacements to drive p-modes to observable amplitudes on a high-gravity white dwarf. The second class of modes represents the g-modes, where gravity is the dominant restoring force. These have timescales of hundreds of seconds and longer, just like the observed oscillations in white dwarf stars. Also, the motions are predominantly horizontal---along gravitational equipotential surfaces---which is energetically more favorable for driving to observable amplitudes. These expressions also indicate that the frequencies of the g-modes decrease with increasing radial overtone number (shorter wavelength) with an accumulation point at zero frequency. We can construct an expression for g-mode frequencies, σ, from the limit for g-mode frequencies by an integration over the star. We then arrive at an expression,

$$\sigma_{k,\ell,m} \approx \left\langle \frac{N^2 \ell(\ell+1)}{k^2 r^2} \right\rangle^{1/2} + \left[ 1 - \frac{C_k}{\ell(\ell+1)} \right] m\Omega \qquad (1)$$

We include the second term on the right-hand side of the equation to indicate the effects of slow rotation on the frequencies. Note that there is a somewhat similar effect due to magnetic fields, which we do not explicitly include here. The effect is to break the spherical symmetry and to make the azimuthal quantum numbers *m* nondegenerate. Here Ω is the rotation frequency, and the constant $C_k$ is a quantity depending on the eigenfunction but approximately equal to one in most cases. This last term produces the potential for fine structure, splitting each radial overtone into $2\ell+1$ components. Observing the spacing between these components then allows us to measure the rotation rate of the star.

The first term in the above equation dominates in the slow rotation, small magnetic field limit. We see another important feature of the g-mode frequencies: In a compositionally homogeneous star, we expect the spacing of consecutive radial overtone g-modes, for a given spherical harmonic degree $\ell$, to be approximately uniform in period. The spacing depends on the integrated average of the Brunt-Väisälä frequency, *N* and so, within an instability strip, is set primarily by the total mass of the star. Deviations from this uniform spacing give information about compositional stratification and allow us to measure the masses of the different layers. The large separation of the instability strips in effective temperature alters the $\langle N^2 \rangle$, explaining different mean period spacings between strips.

We can use plots of the run of the the square of the Brunt-Väisälä, $N^2$, and Lamb, $S_\ell^2$, frequencies through the star to determine where modes of given frequencies will propagate. Such diagnostic plots are known as propagation diagrams. They contain essentially all the information we can obtain through asteroseismological analysis of the pulsation frequencies of the star. The ultimate in asteroseismological analysis is to use the distribution of observed frequencies to determine empirically the propagation diagram for the star. If this is known, the constitutive physics can be deconvolved---at least in principle. This analysis is called seismological inversion and has been used successfully so far on only one star, the Sun. This is because the number of modes observed in other stars is limited. But there is hope that this technique can be applied to white dwarfs in the future---more on this below. For now, we use mostly the forward technique of matching the observed frequencies to models and trying to find the theoretical model that best fits the observed periods.

Figure 2 shows an example of a propagation diagram for a massive white dwarf model. Our dispersion relation shows us that the region of propagation for the g-modes is the region underneath both curves. We notice two things. First, these are global oscillations involving the whole star. Our choice of units for the horizontal axis, radius over pressure, allows us to see features near both the surface and the center (the center is to the left). The vertical axis reveals that the g-modes have periods much longer than 10s of seconds and in the range of 100 to about 1000 seconds, just the timescales observed in Landolt's star and all the other pulsating white dwarf stars. Figure 2 illustrates that the modes penetrate deeply into the interior---any information we gain from the oscillations is information not only about the envelopes, but about the interiors as well. Montgomery, Metcalfe & Winget (2003) show that the core and envelope act to affect period distributions in a symmetric way and point to a way to break this symmetry.

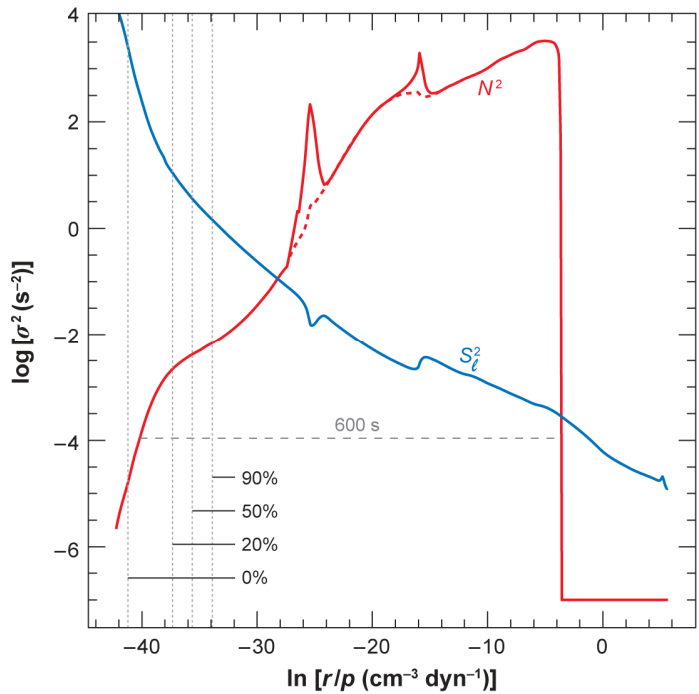

**Figure 2** A propagation diagram for a model of a DAV white dwarf with $1.1 M_\odot$. This illustrates the general behavior of a propagation diagram for a white dwarf model and is appropriate to $BPM\ 37093$. The two solid curves are the Brunt-Väisälä frequency squared, $N^2$, and the square of the Lamb frequency is given by $S_\ell^2$. The vertical dashed lines indicate the location of the crystallized fractional mass. The horizontal axis is the natural log of the ratio of radius to pressure. This gives appropriate resolution at both the center and the surface of the model.

### 3.4. When Frequencies Change

As McGraw et al. (1979a) first pointed out, measurements of the rate-of-period-change in a pulsating white dwarf star constrain many fundamental physical quantities and their time derivatives---providing an illustration of the new precision asteroseismology. As we will demonstrate below, this is a bonanza for physics.

The pulsation periods, P, change slowly with time owing to variations in both temperature, T, and radius, R:

$$\frac{dP}{dt} = a\frac{dT}{dr} + b\frac{dR}{dt}, \qquad (2)$$

where *a* and *b* are constants of order unity ([Winget, Hansen & Van Horn 1983](#)).

To make such a sensitive measurement, all data---often spanning many years, even decades---must be on a uniform time-base. It is not sufficient to use Heliocentric Julian Ephemeris Date. We must transform all data times measurements from Coordinated Universal Time (UTC) to the uniform Barycentric Julian Coordinated Date (TCB) scale, using, for example, the Jet Propulsion Laboratory (JPL) DE405 ephemeris ([Standish 2004](#)) to model Earth's motion about the barycenter of the Solar System.

Next, we must accurately determine the periods and amplitudes in the star. These must be stable enough to bridge, without ambiguity in cycle counts, any gaps in the data. Then we compute an observed time of maximum for each individual night of observation.

The observed times of maximum light are now determined and on a uniform time-base. They are ready to be fit to the equation for the observed minus calculated times of maxima ($O - C$).

If the period is changing slowly with time, we may expand $T_{\max}$ in a Taylor series, discarding terms beyond the quadratic:

$$T_{\max} = T_{\max}\Big|_{E_0} + \frac{dT_{\max}}{dE}\Big|_{E_0}(E - E_0) + \frac{1}{2}\frac{d^2 T_{\max}}{dE^2}\Big|_{E_0}(E - E_0)^2.$$

$E$ is the epoch of the time of maximum, i.e., the integer number of cycles after our first observation. Using $P$ for the period, we can write:

$$\frac{d^2 T_{\max}}{dE^2} = \frac{dP}{dE} = \frac{dt}{dE}\frac{dP}{dt} = P\frac{dP}{dt}$$

and expand the quadratic term, assuming $2E_0 \ll E$, to get:

$$T_{\max} = T_{\max}{}^0 + P \cdot E + \frac{1}{2} P \cdot \dot{P} \cdot E^2.$$

If we next define the observed times of maxima $O \equiv T_{\max}{}^{obs} = T_{\max}$, and the calculated times of maxima $C \equiv T_{\max}{}^1 + P_1 \cdot E$, we get:

$$(O-C) = \Delta E_0 + \Delta P \cdot E + \frac{1}{2} P \cdot \dot{P} \cdot E^2$$

where $\Delta E_0 = (T_{\max}{}^0 - T_{\max}{}^1)$, and $\Delta P = (P - P_1)$.

When we fit a parabola to the ($O-C$) diagram, we obtain the rate of period change for the pulsation, $\dot{P}$, as well as corrections to the period and epoch. We note that at this writing, the fitting of the whole light curve with a term proportional to $\sin\left[\frac{2\pi}{(P+\frac{1}{2}\dot{P})}t + \phi\right]$ by nonlinear least squares gives unreliable uncertainty estimates and the alias space in $P$ and $\dot{P}$ is extremely dense due to the long data time spans typically involved (O'Donoghue 1994).

Of all possible phenomena that may affect $\dot{P}$, cooling has the longest timescale. If the observed $\dot{P}$ is low enough to be consistent with evolution, we can conclude that other processes, perhaps a global magnetic field, convection-induced magnetic fields, or diffusion-induced changes in the boundary layers, are not present at a level sufficient to affect the rate of period change, $\dot{P}$.

**3.4.1. PROPER MOTION** Pajdosz (1995) discusses the influence of the proper motion of the star on the measured $\dot{P}$:

$$\dot{P}_{\mathrm{obs}} = \dot{P}_{\mathrm{evol}}\left(1 + v_r/c\right) + P\dot{v}_r/c ,$$

where $v_r$ is the radial velocity of the star. Assuming $v_r/c \ll 1$ he derived the component of the rate of period change due to the proper motion of the star,

$$\dot{P}_{\mathrm{pm}} = 2.430 \times 10^{-18} P[s]\left(\mu["/yr]\right)^2 d[\mathrm{pc}] ,$$

where $\dot{P}_{\mathrm{pm}}$ is the effect of the proper motion on the rate of period change, $P$ is the pulsation period, μ is the proper motion, and $d$ is the distance.

**3.4.2. CORE COMPOSITION** The heavier the nuclear particles that make up the core of a white dwarf, the faster it cools. The best estimate of mean atomic weight, $A$, of the core comes from the comparison of the observed $\dot{P}$ with values from a sequence of evolutionary models of white dwarfs. For example, Brassard et al. (1992) computed rates of period changes for 800 DAV evolutionary models with various masses, all with carbon cores but with differing He/H surface layer masses. They obtained values similar to those of Bradley & Winget (1991). The average value of $\dot{P}$ for all $\ell = 1$, 2, and 3 modes with periods around 215 s in models with an effective temperature around 13,000 K and a mass of $0.5 M_\odot$, is $\dot{P}(\text{C core}) = (4.3 \pm 0.5) \times 10^{-15}$ s s$^{-1}$. Benvenuto, García-Berro &Isern (2004) C/O models similarly give $\dot{P}(\text{C/O core}) = (3-4) \times 10^{-15}$ s s$^{-1}$. Using a Mestel-like cooling law (Kawaler 1986, Mestel 1952), i.e., $\dot{T} \propto A$, where $A$ is the mean atomic weight in the core, we can write for hot DAVs

$$\dot{P}(A) = (3-4) \times 10^{-15} \frac{A}{14} \text{ s s}^{-1}.$$

There are similar calculations for DOVs (e.g. Althaus et al. 2008a) and DBVs (e.g. Kim, Winget & Montgomery 2006).

**3.4.3. REFLEX MOTION** The presence of an orbital companion could contribute to the observed rate of period change. When a star has an orbiting companion, the variation of its line-of-sight position with time produces a variation in the time of arrival of the pulsation maxima: the light travel time between the star and the observer depends on the reflex motion of the white dwarf around the barycenter of the system. The Doppler shift $\Delta P_{pul} = (P_{pul} - P_{obs})$ on the pulsation period is $\frac{\Delta P_{pul}}{P_{pul}} = \frac{v_r}{c}$, where $v_r$ is the radial velocity of the pulsating star and $c$ is the speed of light. Taking the derivative with respect to time,

$$\dot{P}_{pul} = \frac{dP_{pul}}{dt} = \frac{P_{pul}}{c} \frac{dv_r}{dt}.$$

Defining the inclination angle $i$ as the angle between the plane of the sky and the plane of the orbit and $\theta$ as the position angle in the orbit measured from the line-of-sight direction, the observed radial velocity is then $v_r = v_{orb} \cos\theta \cdot \sin i$. The maximum value of the radial velocity variation occurs for an edge-on system ($i = 90^o$) and for $\theta = 0$ (tangential velocity = 0):

$$\left(\frac{dv_r}{dt}\right)_{max} = v_{orb}\frac{2\pi}{P_{orb}} = \frac{2\pi r_1}{P_{orb}}\frac{2\pi}{P_{orb}}.$$

Here $r_1$ is the distance between the pulsating star and the system center-of-mass, and $P_{orb}$ is the orbital period. Using Kepler's third law, $GM_T = a^3\left(\frac{2\pi}{P_{orb}}\right)^2$, where $a$ is the separation between the two stars, $M_T$ is the total mass of the system, $r_1 = \left(\frac{M_2}{M_T}\right)a_T$, and $M_2$ is the mass of the companion star, we obtain:

$$\dot{P} = \frac{P_{pul}}{c}\frac{GM_2}{a_T^2} = 1.97\times 10^{-11} P_{pul}\frac{M_2/M_\odot}{(a_T/AU)^2}\,\text{s/s}.$$

This allows the detection of companions---planets or stars---orbiting the pulsating white dwarf, as we will discuss in section 9.4.

## 4. SORTING OUT THE PULSATING WHITE DWARF STARS

More extensive histories of the early days of the field than we gave in section 1 have been written by Nather (1978), Winget (1988), Kepler & Bradley (1995), and many others. We reexamine some of landmark results because they still contain puzzles, surprises, and important lessons relevant to future plans.

Four years of work and the discovery of several more pulsating white dwarf stars were required to solve the mystery of Landolt's star. Observations were well ahead of theory. Warner & Robinson (1972) and Chanmugam (1972) independently discovered that these timescales for periodic variability were consistent with nonradial gravity mode pulsations. For the g-modes, the horizontal displacements are larger by a factor of about 1000 than the radial displacements. With hindsight, this is what you expect for self-excited pulsations in high-gravity objects: movement along gravitational equipotential surfaces is strongly favored energetically. Warner & Robinson's work had a profound impact on everyone's thinking about pulsating white dwarf stars, and most importantly, on Carl Hansen & Yoji Osaki. Osaki & Hansen (1973) numerically solved nonradial oscillation equations in the quasi-adiabatic approximation for homogeneous carbon white dwarf models. This work demonstrated not only that nonradial modes have the correct periods but also that the low-radial overtone g-modes have about the right distribution to explain the multiple modes observed. Then (as now), no specific selection mechanism was obvious that would favor one radial overtone over another, or one spherical harmonic degree, $\ell$, over another, but the quantitative agreement of the observed periods with periods in the models was good.

In the years following the discoveries of Landolt (1968) and Lasker & Hesser (1969, 1971), there were an abundance of reports of high-frequency variability in white dwarf stars of all colors and types, as described in McGraw (1977, 1979). They provided a profound misdirection for theoretical explorations of the cause of the variations: It appeared the surface composition and temperature had nothing to do with the excitation of the observed pulsations. This observational misstep set back the field by years. Gerard Vauclair (1971a,b) had demonstrated that radial pulsation instabilities in models of DA white dwarfs occur near the onset of hydrogen partial ionization at ~10,000 K. Then, after Warner & Robinson (1972) pointed out that the observed periods were consistent with nonradial g-modes. Connecting these results would have been straightforward.

If not surface composition or temperature, what caused the pulsations? The cores were presumably very similar. Why some stars pulsed while others did not remained a mystery.

Breaking this logjam required a revolution in instrumentation. Ed Nather provided just such a revolution with the development of the two-star, time-series photometer (Nather 1973). This instrument bore the Nather trademark: It was science-driven and simple. It did one thing, and it did it very, very, well. His photometers revolutionized observations of compact variable stars. The instrument alone was not enough; the systematic application of it by John McGraw, Rob Robinson, and Ed Nather was necessary, too.

Simultaneous light curve of the target and comparison star with the two-star photometer allowed Nather and collaborators to separate variations in atmospheric transparency from variations in intrinsic stellar brightness. Prior to this, relative photometry was a difficult art. Limitations of the human eye prevent detecting variations smaller than of order 10%. This made it difficult to judge when it was truly photometric, particularly so at new moon when most observations of white

dwarfs---relatively faint objects---are made. Slow transparency variations can also occur on the same timescales with the result that nonvariables were sometimes reported as variables. Additionally, this effect complicated the accurate measurement of the phases and amplitudes of pulsations in real variables.

Armed with Nather's two-star photometer, Robinson, Nather & McGraw (1976) systematically reobserved the claimed pulsators, carefully examining them for intrinsic variations under the photometric skies of the McDonald Observatory. One by one the nonvariables fell away, and gradually new ones were discovered. Nather (1978) provided another key piece of the solution to the puzzle. He showed that the single white dwarf pulsators must be considered a separate class of objects from white dwarf stars in interacting binaries. A clear picture emerged: Only white dwarf stars of spectral type DA were observed to pulsate, and they did so only in a narrow range of colors. Robinson (1979) reviewed the observational and theoretical evidence and demonstrated that the DAV (or ZZ Ceti stars) are a very homogeneous class of otherwise normal DA white dwarf stars pulsating in nonradial g-modes. Temperature variations are the dominant effect in producing the observed light variations. McGraw (1977, 1979) brought us to what is essentially our current understanding with the realization that the colors corresponded to temperatures near the H-opacity maximum, indicating that the partial ionization zone of hydrogen was responsible for driving the pulsations. The collective impact of this important work went through the astronomical community like a shock wave, precipitating discovery everywhere it touched.

Theoretical successes now followed one another in rapid succession. First, Wojtek Dziembowski & Detlev Koester (1981) found instabilities in nonradial g-modes, but the driving was in the underlying He layer, not in the surface H layer. Noel Dolez & Gerard Vauclair (1981) were the first to find driving of nonradial g-mode pulsations in DA white dwarf models, followed nearly simultaneously by others (Winget, Van Horn & Hansen 1981; Starrfield et al. 1982; Winget et al. 1982a). Theorists were finally able to find in their models the association of the excitation by the zone of partial H ionization discovered by McGraw.

The demonstration of driving from the H-partial-ionization zone led Winget (1981) and Winget et al. (1982a) to investigate models of DB white dwarf stars for possible instabilities owing to the surface He partial ionization at a correspondingly higher temperature. They found instabilities in their models and predicted pulsations in DB white dwarf stars near the He I opacity maximum associated with the onset of significant partial ionization.

Observations soon caught up. A systematic survey of the DB white dwarf stars demonstrated that the brightest DB with the broadest He I lines, GD 358, did indeed pulsate in nonradial g-modes---remarkably similar to the large amplitude DAV pulsators (Winget et al. 1982b).

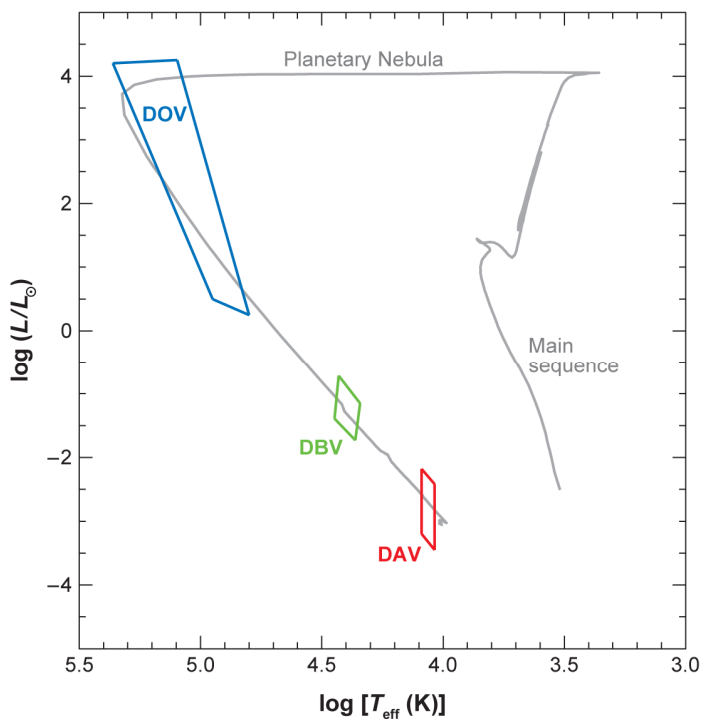

**Figure 3 A 13-Gyr isochrone with z = 0.019 from Marigo et al. (2007), on which we have drawn the observed locations of the instability strips, following the nonadiabatic calculations of Córsico, Althaus & Miller Bertolami (2006) for the DOVs, the pure He fits to the observations of Beauchamp et al. (1999) for the DBVs, and the observations of Gianninas, Bergeron & Fontaine (2006) and Castanheira et al. (2007, and references therein) for the DAVs.**

The observed pulsating white dwarf stars lie in three strips in the H-R diagram, as indicated in Figure 3. The pulsating pre-white dwarf PG 1159 stars, the DOVs, around 75,000 K to 170,000 K have the highest number of detected modes. The first class of pulsating stars to be predicted theoretically before discovery, the DBVs are found around 22,000 K to 29,000 K, and the DAVs around 10,850 K to 12,270 K. The 172 pulsating white dwarf stars known in March 2008 are all in the thin disk of our Galaxy, because they are intrinsically faint. Given the implied space-density,they are easily the most numerous class of variable stars (Cox 1980). Their structure is simple; seismology, therefore, gives structural information with detail and precision. Because of their high densities and internal temperatures, they are tools to study physics at high energies and densities, where quantum effects are dominant and post-Newtonian corrections are (still) not.

Identification of the spherical harmonic degree, $\ell$, is possible using multiplets for some pulsating DOVs and DBVs. Alternatively, mode identification is possible through the technique of chromatic amplitudes changes from UV to optical (Robinson et al. 1995, Kepler et al. 2000a, Castanheira et al. 2004, 2005), line profile variations (Clemens, van Kerkwijk & Wu 2000; Thompson et al. (2003); Kotak, van Kerkwijk & Clemens (2004) used the ratio of the amplitudes of combination modes to their parent modes for DBVs and DAVs to identify the spherical harmonic degree, $\ell$, of the parent modes. Nearly all the modes identified up to today have $\ell = 1$ or 2. Yeates et al. (2005) propose to use the amplitudes of the linear combination peaks caused by distortions in the subsurface convection layer to identify $\ell$ using the amplitude equations of Wu (2001).

As we were preparing this review, a fourth instability strip may have been discovered. Montgomery et al. (2008) predict pulsations resulting from carbon partial ionization driving. and report the discovery of a new variable star, SDSS J142625.71+575218.3, that fits this description. It is a hot carbon-atmosphere white dwarf star (hot DQ star), discovered by Dufour et al. (2007). These stars were predicted to pulsate in a narrow instability strip just below 20,000 K by Montgomery as the result of carbon parial ionization. The variable star has a regular period of 418 seconds. Its measured effective temperature is just under 20,000 K and it has a high surface gravity (Dufour et al. 2007).

These properties place it within the theoretical instability strip of carbon partial ionization (Montgomery 2008). Fontaine et al. (2008) have also published a predicted theoretical instability strip that may be appropriate for these stars, but their models predict a different mechanism. These models are unstable from He-partial ionization. This requires a helium abundance of about 50%, by mass, or greater. Observations of the He-abundance in this star will be critical in testing the two theories. We remain tentative about the identification of the star as a pulsator only because the pulse shape of the observed variations resembles the pulse shape of an interacting He-surface composition interacting binary white dwarf, AM CVn, more than any other pulsating white dwarf star. Beyond the pulse shape, there is no other evidence that this star is a binary and a carbon analog to the interacting helium twin-degenerates. Such a model seems unlikely and presents other difficulties as discussed by Montgomery et al. (2008). We await future observations before we can completely rule out this possibility. It remains a very exiting object either way.

## 5. CURRENT TOOLS AND IDEAS

### 5.1. High Resolution Power Spectroscopy and the Whole Earth Telescope

By the mid-1980s, it was clear that future progress in extracting physical information from pulsating white dwarf stars required precise identification of the normal mode frequencies in each star. This was problematic, given the multiperiodic nature of white dwarf pulsators. For normal mode analysis, more modes mean more information, but the power spectrum is harder to untangle.

The timescale of 70--1500 s for the pulsations is convenient for observations of a few hours with the star near the meridian. One can gather many data points per cycle, and tens to hundreds of cycles of data. But the pulsation modes in a complex multiperiodic pulsator are difficult to unravel with data sequences of 2--4 hr. The reason is that the spacing of the radial overtones and the possible frequency splittings must all be resolved (see Equation 1). These splittings are hours for typical consecutive radial overtones and days for typical rotation periods. The problem is exacerbated by the possibility that dipole and quadrupole modes are simultaneously present. Without the ability to disentangle these modes, we cannot account for the observed variations in amplitude and frequency.

In some objects frequencies come and go, dropping below detectability. Thinking in terms of the Fourier transform of the light curves, we were facing a frequency resolution problem. We were unable to understand the observed variations in the power spectra of many pulsating white dwarf stars from night to night. Which were attributable to real instabilities, mode interactions, and nonlinear effects, and which were due to beating of closely spaced modes? Real asteroseismology requires we resolve these issues. The resolution of a light curve is set by its length; we needed longer uninterrupted light curves.

Our first response to the demand for resolution was to push individual runs to the maximum length possible, going as nearly as possible from horizon to horizon. Even with this, light curves of eight hours were clearly not enough. In response, observing campaigns were organized using sites separated in longitude. The first attempt involved Cerro Tololo Interamerican Observatory in Chile and the McDonald Observatory in Texas (e.g., [Stover et al. 1980](#)). Later we added Darragh O'Donoghue at SAAO in South Africa. With this approach, we made the first direct detection of stellar evolution in the DOV star, PG 1159-035 ([Winget et al. 1985](#)). But even this was not enough.

Theory was just catching up with Steve Kawaler's analysis of the DOV prototype star, PG 1159-035 ([Kawaler 1986](#)). Based on his analysis of the two principal modes observed, he predicted the existence of a number of other modes. But finding them was no easy observational matter---two sites were not enough. We needed a dramatic improvement in the resolution of the light curves. Theory had gone as far as it could. Ed Nather found a way to eliminate the regular gaps in the data caused by the inevitable regular appearance at each site of the Sun. Ed's elegant idea was to use optical observatories spread around the globe to follow the observations of a single star westward with the darkness, observing continuously through an entire lunar dark-time. He later dubbed his vision the Whole Earth Telescope (WET). Nather pointed out that it is the inverse of the old British Empire, "With the entire planet as its mount, the Sun never rises on the Whole Earth Telescope."

The WET was the first of several global networks for asteroseismology to come on-line. It did so by virtue of its organization. Reflecting the spirit of its founders, it ignored political and bureaucratic entities, allowing what eventually became a group of more than

50 scientists from more than 14 countries to organize themselves based on common scientific interest. The detailed history of the development of this grand astronomical instrument and its first scientific applications are described in detail by Nather et al. (1990), [Winget et al. (1991)](#), [Winget et al. (1994)](#).

The WET brought a revolution with nearly uninterrupted light curves spanning one to two weeks. The data density and duration gave birth to "High signal-to-noise, high-resolution, power spectroscopy," as Nather puts it.

New results and new steps forward followed, surviving two changes in leadership and locale. After its first decade the WET headquarters moved from Texas to Iowa State University, into the hands of Steve Kawaler and Darragh O'Donoghue, along with Chris Clemens and Scot Kleinman among others. It is now starting its third decade, as a part of the Delaware Astronomical Research Corporation (DARC) in the hands of Harry Shipman and director Judi Provencal, with the assistance of Susan Thompson and Mike Montgomery.

Just now the WET is being used for an entirely new purpose, to explore the shape of the light curves themselves. We discuss the new science coming from this in Section 8.

**5.2. Charge-Coupled Device Time Series Photometry**

Currently, time series instrumentation has undergone another revolution from employing CCD cameras as detectors, replacing photomultiplier tubes. This gives up a large portion of the high time resolution because of the limitations on readout times. For observations of pulsating white dwarf stars, exposure times longer than a second are most useful, so the increased sensitivity of CCD observations is worth giving up some of the time resolution in most applications.

The prime example of a CCD system is the one conceived by Nather with the help of Anjum Mukadam: Argos ([Nather & Mukadam 2004](#)). The CCD camera is deployed at prime focus on the 2.1-m telescope at the McDonald Observatory, with no corrective optics.

The Argos CCD camera and control system realize an overall improvement in sensitivity of about a factor of nine compared with our older time-series instruments based on photomultiplier tube detectors. This allows observations of many faint white dwarf stars. In addition, by using nearby comparison stars in the field of view, we are able to correct for the effects of thin clouds and nonphotometric conditions, doubling the number of usable nights. All of this is of prime importance, because modern surveys are discovering thousands of previously unknown, faint white dwarfs. For example, using the Argos CCD camera to observe a sample of white dwarfs culled from the Sloan Digital Sky Survey (SDSS) has led to a more than doubling of the previously known population of pulsating white dwarfs (Mukadam et al. 2004a, Mullally et al. 2005). In addition, they have found empirically that the increased timing accuracy of this instrument---in the sense of time tagging each exposure---allows them to determine the baseline pulsation periods more quickly in individual objects. Mullally et al. (2008) calibrated the effect of the timing accuracy using the well-studied DAV G 117-B15A. After three seasons of observations with Argos on the 2.1-m telescope, the period accuracy was equivalent to that obtained in 30 years of archival photomultiplier tube data.

**5.3. Understanding the Amplification of the Pulsation Modes**

Excitation or driving of stellar pulsation modes is really tantamount to amplification. Physically, we envision ever-present and infinitesimal random fluctuations amplified to finite size when driving exceeds damping. Note that we are not quite physically consistent in this picture, as we are limited, for the most part, to linear calculations: growth to finite amplitude violates our approximation. The calculations work pretty well for the adiabatic determination of the normal frequencies.

The association of excitation with partial ionization of the most abundant surface element seems very secure in the DAV, DBV, and probably DOV stars. The theoretical prediction and subsequent detection of pulsations in DB white dwarf stars hinged on this point. Locally in a star, the condition for driving is that the maximum pressure occurs after the maximum density during a pulsation. Further, this condition must occur in a region of the star around the adiabatic/nonadiabatic transition zone. This, very roughly, is the point in the star where the local thermal timescale is of order the pulsation timescale. Much deeper than this, the conditions are adiabatic on the timescale of the pulsations. Far above this point, any perturbation to the energy is quickly carried away---the conditions are strongly nonadiabatic. The net driving or damping is determined by the integrated contributions over the whole star. Clearly, regions far from the adiabatic/nonadiabatic transition zone cannot contribute to the energy integral. If driving outweighs damping, the mode is overstable and should grow in amplitude. The bulk of the action occurs at the base of the partial ionization zone, for modes with periods near the local thermal timescale.

The specific way in which the partial ionization zone drives the pulsation is not as clear. Currently, two different descriptions of the driving have been used. One seems more physical than the other, but neither agree well with crucial observational data.

The first method for computing excitation was used by all researchers mentioned in Section 4. This involves the critical assumption that it is only the radiative luminosity that is perturbed---the convective luminosity is frozen in on the timescale of the pulsations. These calculations involve the $\kappa-\gamma$ mechanism (see, e.g., [Cox 1980](), [Unno et al. 1989]() for detailed discussions). They give instabilities in roughly the regions observed, providing effective temperatures for the hot end of the instability strip, the blue-edge, that are reasonable, qualitatively. The peak growth rates also imply increasing periods with decreasing surface temperature, consistent with the observations.

This was better agreement, if only qualitatively, than anyone had a right to suspect. The reason was well known: The convective turnover timescale is less than or of order the dynamical timescale---i.e., seconds---so convection is not frozen in; it is actually in the opposite limit, adapting nearly instantaneously to the pulsations.

This unsatisfactory state of affairs was first addressed by Anthony (John) Brickhill (1983), and later in more detail in [Brickhill (1990)](), and his subsequent papers (see [Goldreich &Wu 1999](), and references therein). Brickhill's calculations assumed the opposite limit from all previous calculations: instantaneous adjustment of the convective flux to the pulsations---a much more physically self-consistent limit for white dwarf stars. Brickhill was well ahead of his time; his work was not properly appreciated until Peter [Goldreich & Yanqin Wu (1999)]() carried out quasi-adiabatic calculations in the same approximation. Brickhill called this new type of driving convective driving.

These calculations account for the excitation of the modes---the tendency to longer periods at lower effective temperatures---as did the earlier calculations. But they go much farther. Convective driving also explains the lower effective temperature boundary, the red edge, to the pulsation instability strip, and accounts for the pulse shapes of the large amplitude pulsators. This gives us a powerful tool for tracking the behavior of the depth of the convection zone during a pulsation cycle.

We close this section with some problems. It remains to be demonstrated that the Brickhill model explains the light curves of the pulsators near the blue edge. These are qualitatively different from the large-amplitude pulsators, and the models indicate extremely thin surface convection layers. Further, these pulse shapes are the same as the ones we find in the DOV stars, where convection is apparently not present (e.g., Costa et al. 2008) and where the driving seems most likely due to the operation of the classical $\kappa - \gamma$ mechanism.

The waters are further muddied by the recent, somewhat puzzling, results of Agnes Kim et al. (2006). She was searching for a way independent of spectroscopy to improve determinations of effective temperature. This, of necessity, involves looking at the nonadiabatic driving effects. We note that the straightforward evolutionary changes in periods cause individual periods to change by less than of order 10% across the instability strip, not the factor of ten increase observed. So the observed changes must be due to the evolutionary change in the equilibrium structure of the driving region. Kim et al. (2006) idea was to use the observed trends of increasing pulsation periods with decreasing temperature as a relative temperature index following the well-established empirical results of Mukadam et al. (2006, and references therein). They determined a promising empirical result. The problems arise on comparison with theoretical models, assuming a relation between the thermal timescale in the driving region and periods of the excited mode as the star evolves. They found that the observed periods change more slowly with effective temperatures than predicted by either model of driving. Vexing as the assumption of frozen-in convective flux is, the best match was with the most efficient model of convection and the $\kappa - \gamma$ mechanism. The inability of models to explain the change in periods with decreasing effective temperature in a realistic quantitative way is disturbing. Models disagree with observations by a factor of two or more in the exponent. Although convective driving appears not to work well in explaining the evolutionary trend through the strip---we caution that there may be straightforward reasons this is so--- it does an excellent job of explaining the change in the convection zone during the pulsation cycle of an individual large amplitude star, as is discussed in Section 8.

Searching for the underlying explanations for both of these facts should be a fruitful area for exploration in the near future.

## 6. THE STATE OF ASTEROSEISMOLOGY, STRIP BY STRIP

### 6.1. DOVs

The instability strip of the DOVs, a.k.a. pulsating PG 1159 stars, or GW Vir stars, after the prototype PG 1159-035 discovered by McGraw et al. 1979a,b, is around $T_{\text{eff}} \simeq 170,000$ K to 75,000 K and $\log g = 5.7$ to 7.5. The strip includes both the stars without evidence of surrounding planetary nebulae (McGraw 1979, Bond et al. 1984), and the PNNs (Grauer & Bond 1984). Both present detectable evidence of ongoing mass loss. Their atmospheres are mainly composed of He, C, and O, and the pulsators also have strong lines of N (Dreizler 1998). These H-deficient stars may be the evolutionary remnants of stars that have experienced a late He thermal pulse after the star has left the AGB (Althaus, Córsico & Miller Bertolami 2007, and references therein). There are at least 11 pulsators known as of March 2008, and their periods change slowly with time owing to evolutionary changes in both temperature and radius (see Equation 2), with temperature the dominant affect in all stars but the hotter PG 1159 stars and the PNNs (Winget, Hansen & Van Horn 1983; Winget et al. 1985; Kawaler 1986; Costa, Kepler & Winget 1999). The pulsation periods range from 400 to 3000 s---longer for the PNNs (Vauclair, Solheim & Østensen 2005). Pulsations in the prototype, PG 1159-035, have been detected even in X-rays (Barstow et al. 1986). The period spacings for the hotter variables are close to those predicted by asymptotic theory, as they are high-$k$ pulsators, but Córsico et al. (2007) show that there are measurable differences from the asymptotic value, especially for the coolest stars. Presently, the largest uncertainty in the mass determination from the period spacings comes from uncertainties in the theoretical models (Kawaler et al. 1995, 2004; Althaus et al. 2007), not from observational uncertainties (Córsico et al. 2007, Fu et al. 2007, Costa et al. 2008). So an effort in accurate modeling is necessary and underway(Miller, Bertolami & Althaus 2006, Córsico et al. 2007, 2008, Althaus et al. 2008a). The accuracy in the mass determination from the period spacings, even with the uncertainty in the models, is of the order of $\Delta M \simeq 0.02 M_\odot$ (Costa et al. 2003, Córsico et al. 2008; Costa et al. 2008, Althaus et al. 2008b). This is at least an order of magnitude more accurate than the determinations from spectral fitting. As convection is negligible in these stars, the $\kappa-\gamma$ mechanisms at the C and O partial ionization zones are the main drivers, as originally proposed by Starrfield et al. (1983) and confirmed by Bradley (1996), and more accurately confirmed with the evolutionary models of Quirion, Fontaine & Brassard (2004, 2005, 2007), Córsico & Althaus (2005, 2006), and Córsico, Althaus & Miller Bertolami (2006). There are pulsators and nonpulsators in the same region of the **Hertzsprung-Russell** diagram, but there are significant abundance differences. Why some stars pulsate and others do not is an important question for the future.

### 6.2. DBVs

The class of pulsating DB stars, also called V 777 Her stars after their prototype GD 358 (a.k.a. V 777 Her), discovered by Winget et al. (1982b), have atmospheres of helium. There are now 18 pulsators known (Nitta et al. 2007, 2008). The instability strip is located at $T_\text{eff} \simeq 29,000$ K to 22,000 K, with an uncertainty of ~2000 K owing to uncertainties in the temperature determination from spectral fitting and possible contamination by an undetectable amount of H (Beauchamp et al. 1999, Castanheira et al. 2006). The excitation near the blue edge is consistent with the $\kappa-\gamma$ mechanism in the He partial ionization zone, as proposed by Winget et al. (1982b). The DBVs comprise the first class of variable stars predicted theoretically. At the blue edge of the instability strip, the convection zone in the models is so high up in the atmosphere, at $M \simeq 10^{-14} M_*$, and with a thermal timescale around 1 s, that it seems it cannot drive pulsations (Sullivan et al. 2007, 2008). The pulsation spectra in general show a large number of harmonics and combination periodicities, consistent with a thick convection zone distorting the eigenmodes at its base (Ising & Koester 2001; Montgomery 2005a, 2006).

The prototype and brightest known member, GD 358, shows hundreds of linear combination peaks in the Fourier transform of the light curve and strong amplitude changes on timescales of weeks and months (see Figures 4 and 5 and Winget et al. 1994, Vuille 2000, Kepler et al. 2003).

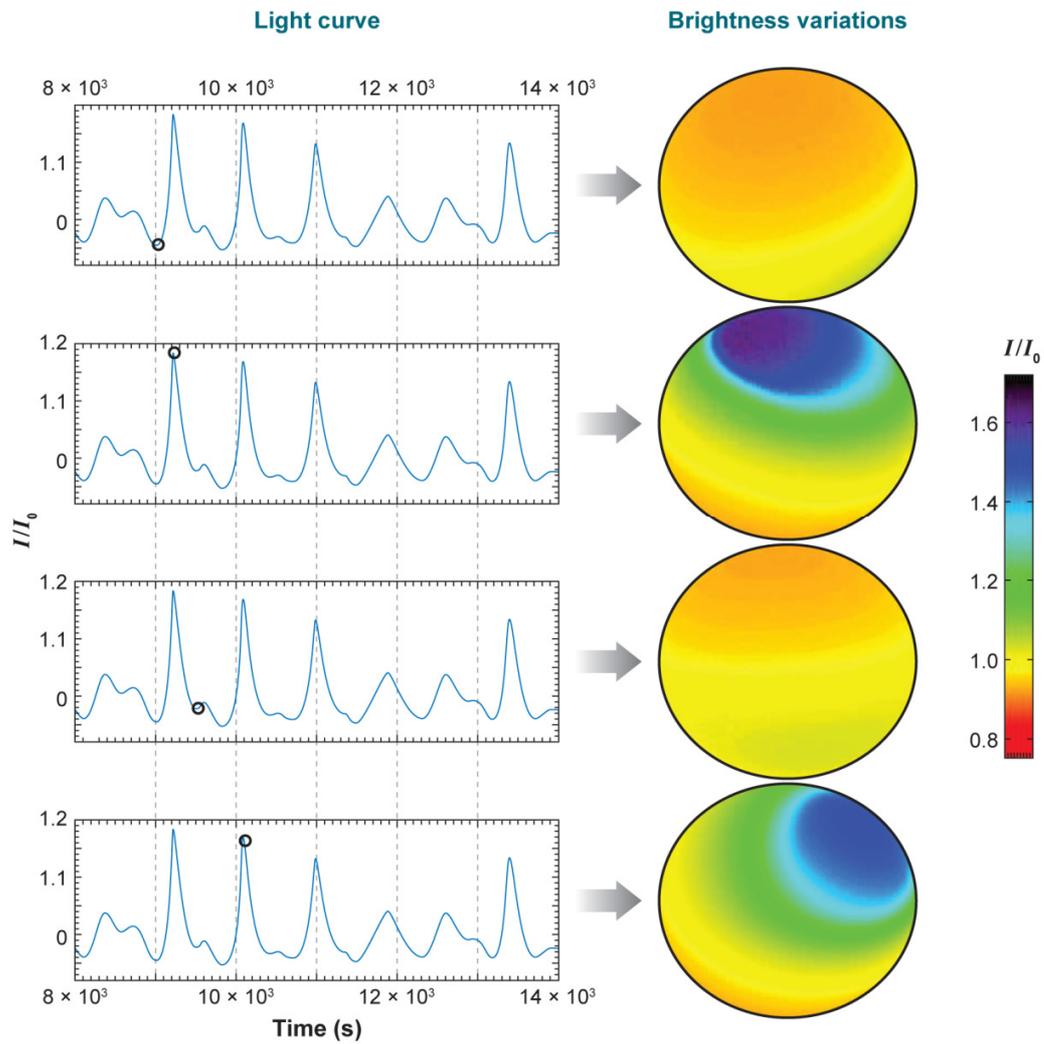

**Figure 4 Surface brightness changes for the DBV GD 358, according to the non-linear convection/pulsation models of Montgomery (2007). The left side shows the position in the observed light (flux) corresponding to the surface brightness changes modelled on the right (intensity).**

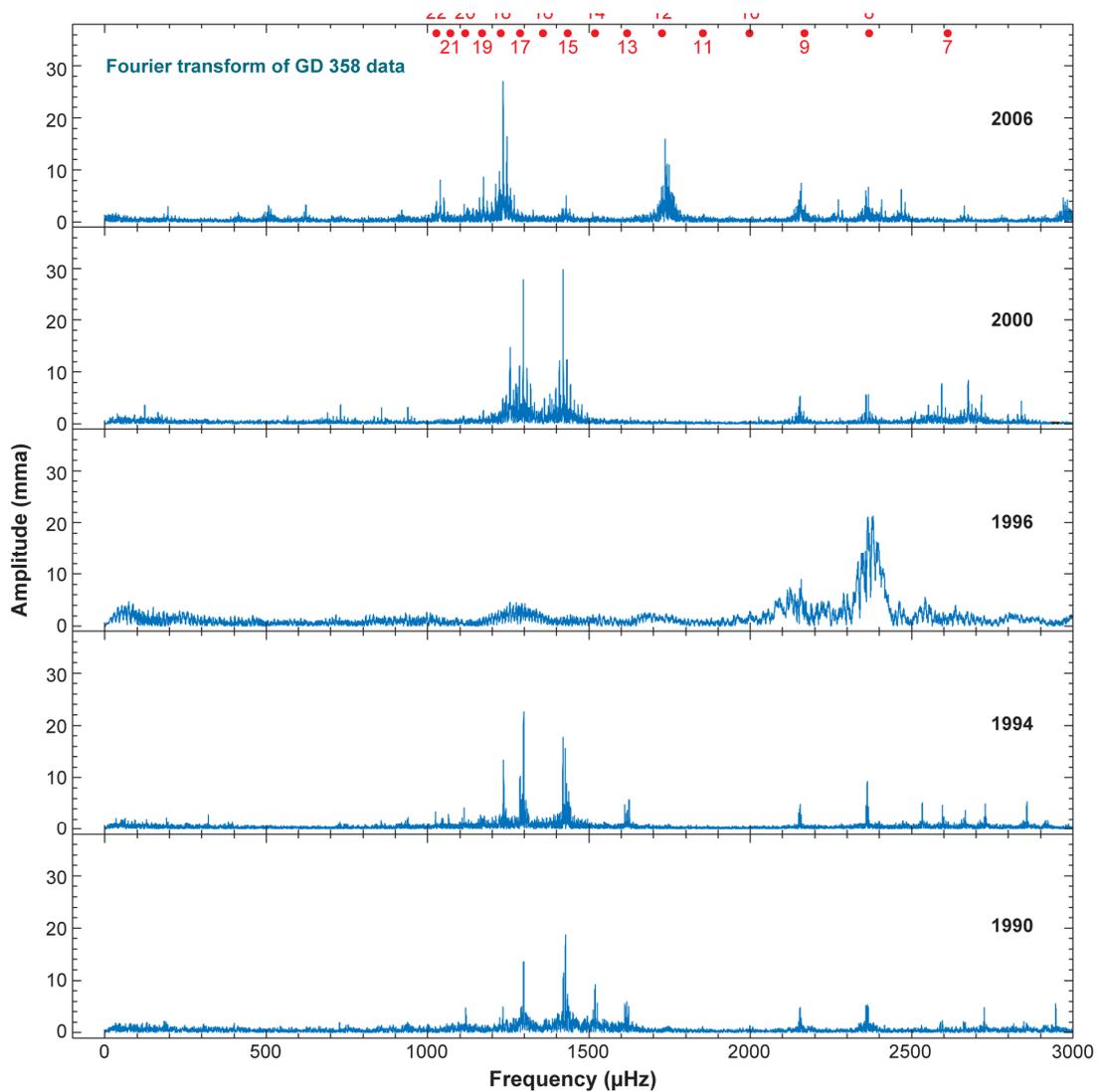

**Figure 5** Fourier transforms of the light curves of GD 358 of data taken on several Whole Earth Telescope (WET) campaigns. The red marks on the top indicate the expected location of modes following an asymptotic $\Delta P$. Even though there are still remaining small alias peaks introduced by the existence of some gaps in the light curves, there are clear changes in amplitude of the excited modes.

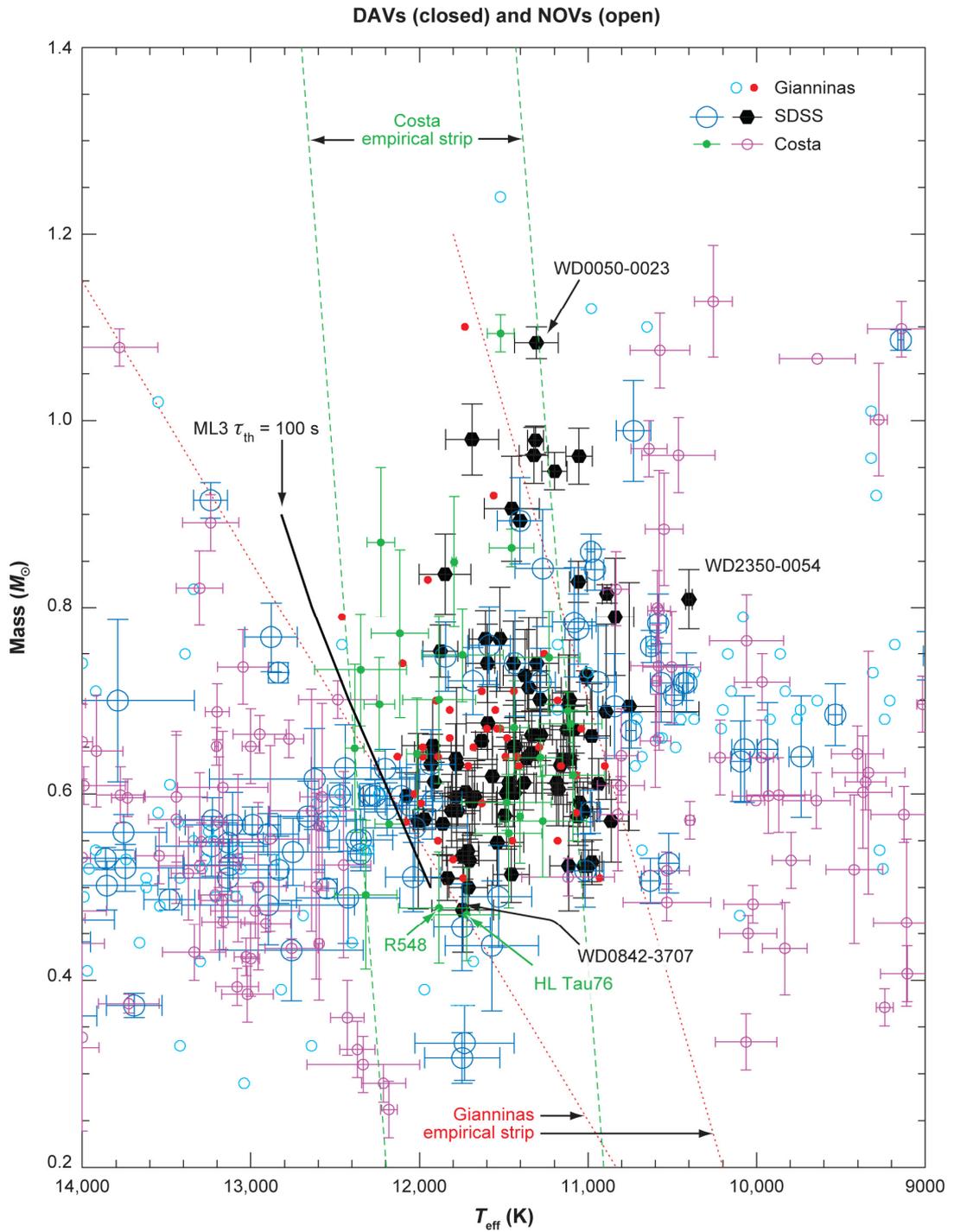

**Figure 6** Location of the DAV instability strip, showing the variables and nonvariables, both from the bright sample (Giannias, Bergeron & Fontaine 2006, Costa 2007) and the Sloan Digital Sky Survey (SDSS) sample (Kepler et al. 2007). The variables are plotted with filled symbols and, for those for which no pulsation

was detected up to the available detection limit, open symbols. The symbols without error bars are from Giannias, Bergeron & Fontaine (2006) because they only cite an average internal uncertainty around 200 K. The dashed lines are only empirical lines drawn to include all the bright variables. The $\tau_{th}$ ~100s solid line corresponds to the theoretical models with ML3 convection theory for which the timescale at the base of the convection zone is 100s.

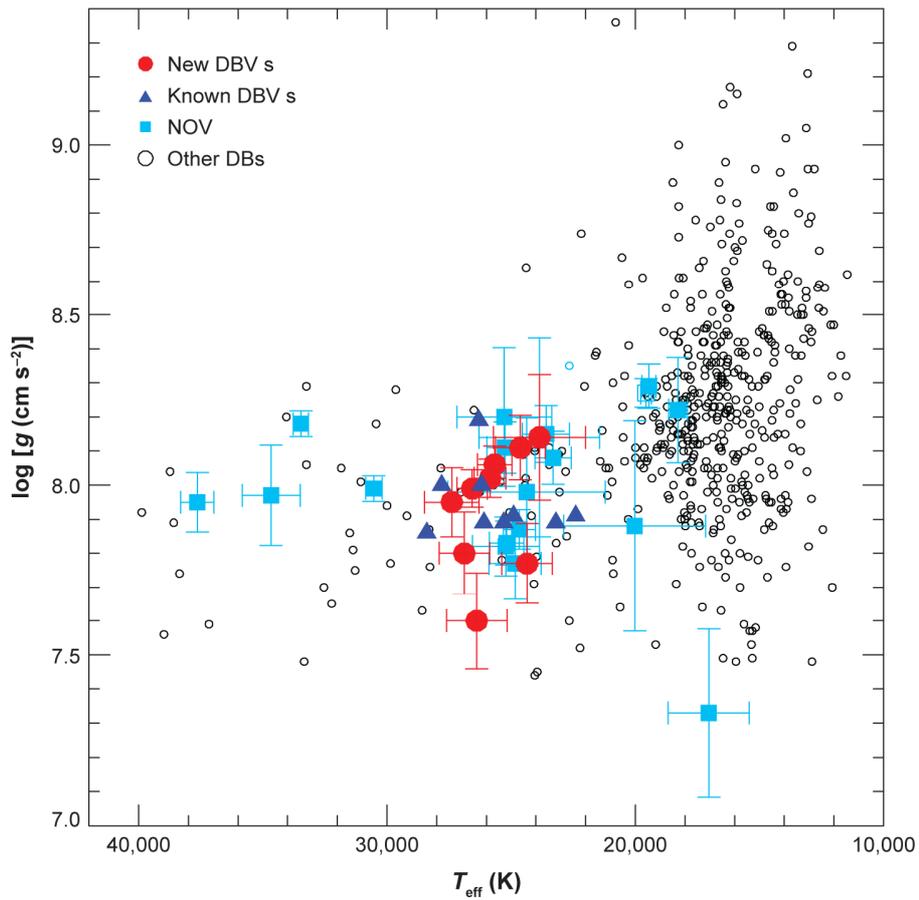

**Figure 7** Location of the DBV instability strip, by Nitta et al. (2008). The effective temperatures and gravities are those determined from models of pure He by Beauchamp et al. (1999) for the bright sample and Eisenstein et al. (2006) for the SDSS stars. The open symbols represent DBs for which no time series photometry has been obtained yet.

The periods range from 140 s to around 1000 s and the uncertainties in temperatures, coupled with the possible contamination of a small amount of H, if any, in the spectra of a few DBs, make the analysis of the purity of the DB instability strip difficult (see Figure 7).

Brassard & Fontaine (1999) determined that nonadiabatic calculations with an ML1/$\alpha = 1$ description of convection better match the observed blue edge, whereas ML2/$\alpha = 1.25$ better fits the synthetic atmospheric spectra (Beauchamp et al. 1999). Gautschy & Althaus (2002) studied the theoretical instability strip with self-consistent diffusion, Canuto, Goldman and Mazzitelli (1996) CGM convection theory, and OPAL opacities, finding a blue edge roughly in agreement with the observations, but a red edge a few thousand degrees too cool, even taking into account leakage of pulsations into the atmosphere. Their mode trapping analysis suggested a few islands of modes separated by a few hundred seconds of stable modes, similar to the observations of GD 358 in Figure 5.

Opal and OP opacities are the calculations by the group at the Lawrence Livermore National Lab, available at http://physci.llnl.gov/Research/OPAL/opal.html and the Opacity Project, an international collaboration, available at http://vizier.u-strasbg.fr/topbase/op.html.

### 6.3. DAVs

The DAVs or ZZ Ceti class of pulsating white dwarf stars has 143 known members at this writing (March 2008) (Gianninas, Bergeron & Fontaine 2006; Castanheira et al. 2007, and references therein). It was the first class to be observed, where our story began with Landolt (1968) and the mysterious variations of HL Tau76. Soon afterward, Lasker & Hesser (1969) found G44-32, with periods around 600 and 820 s, followed by R 548 = ZZ Ceti, with periods of 213 s and 271 s (Lasker & Hesser 1971). The class was first studied by McGraw & Robinson (1976). Robinson, Nather & McGraw (1976) first detected rotational splittings in R 548, and McGraw (1979) and Robinson, Kepler & Nather (1982) showed that the light variations were dominated by changes in temperatures caused by *g*-mode pulsations. A possible filter mechanism that selects which modes get excited to observable amplitudes, mode trapping, was studied by Winget, Van Horn & Hansen (1981) and Córsico et al. (2002), and references therein). The exact mode selection mechanism remains a mystery. Some pulsators have small amplitudes and sinusoidal light curves (e.g.,

Kepler et al. 1982), others have high amplitudes and many harmonics and linear combination peaks (e.g., Vuille 2000).

Recent calculations of models using OP and OPAL opacities indicate that the superficial convection zone is carrying about 90% of the flux even at the blue edge. But the current models still have difficulties showing a significant convection layer at the observed blue edge of the instability strip without extreme assumptions for the convective efficiency.

Gautschy, Ludwig & Freytag (1996) included convection-pulsation interaction, and Wu (2001) found a red edge around $P \simeq 1400$ s owing to this interaction. Even the leakage of pulsation energy into the atmosphere as studied by Hansen, Winget & Kawaler (1985) for DAVs and Gautschy & Althaus (2002) for DBVs cannot explain the observed red edges.

The rotation periods derived from pulsation splittings are all ~1 day, consistent with those observed by line broadening. This implies that even nonmagnetic white dwarf stars are slow rotators. The velocity fields in line profiles are starting to be detected with time-resolved spectra taken at the Keck 10-m telescopes and with the VLT (Koester & Kompa 2007).

## 7. A POTPOURRI OF PUZZLES

### 7.1. Purity of the DAV Instability Strip

The investigation of the purity of the ZZ Ceti instability strip clearly depends on the observed amplitude sensitivity in any search for pulsations. The investigation also depends on the accuracy of the determination of the effective temperature and gravities, as the instability strip ranges only around 1500 K in $T_{\text{eff}}$ and depends on gravity. With the high signal-to-noise (S/N) spectra for the bright sample, Bergeron et al. (1995, 2004) and Giannias, Bergeron & Fontaine (2005, 2006) find a pure instability strip---there are around 20 stars for which no pulsation was observed inside the same instability strip if one uses the less accurate determinations of surface parameters for the fainter SDSS variables and the relatively high detection limits of Mukadam et al. (2004a) and Mullally et al. (2005), of order 4 mma.

Castanheira et al. (2007) found low-amplitude variability for two stars reported as nonvariables in the aforementioned searches, and Kepler et al. (2006) find that the uncertainties in the SDSS parameters are a substantial fraction of the width of the instability strip. Figure 6 shows the current status of the DAV instability strip. Mukadam et al. (2006) suggest that we can use the observed pulsation periods to determine $T_{\text{eff}}$, as there is a strong correlation between period and $T_{\text{eff}}$. Even with the small number of pulsations detected in the DAVs, seismology indicates that $M_H \simeq 10^{-4}$ to $10^{-9.5} M_*$ with an average of $M_H \simeq 10^{-6.5} M_*$ (Castanheira & Kepler 2008). This is an important limit in the study of the chemical evolution of the surface composition of white dwarf stars owing to ongoing diffusion, radiative levitation, and convection. With this relatively thick hydrogen layer, even though it is thinner than the canonical value of $M_H \simeq 10^{-4} M_*$ from current evolutionary models, convection still cannot bring subsurface He to the photosphere, increasing the pressure broadening (e.g., Bergeron, Gianninas & Boudreault 2007).

Starting with Bergeron et al. (1995), high signal-to-noise (S/N $> 60$) optical spectra fitted with synthetic spectra derived from models atmospheres have been used to measure $T_{\text{eff}}$ and $\log g$ of variables and nonvariables around the DAV instability strip. There is a small dependency on the mass of the star, determined from comparison of atmospheric parameters to evolutionary models (e.g., Wood 1995, Althaus et al. 2005), both theoretically (e.g., Bradley 1996) and observationally (Giovannini et al. 1998; Gianninas, Bergeron & Fontaine 2006). The uncertainties in the temperature determinations have always been greater than 300 K, if we compare determinations from different spectra of the same star, which is a substantial fraction of the 1500-K width of the strip.

Of the 143 DAVs known in March 2008, 83 were discovered in the SDSS using temperatures determined from fits to S/N $< 30$ SDSS spectra. SDSS. Mukadam et al. (2004b) and Mullally et al. (2005) found several stars with temperatures inside this low S/N instability strip that did not show detectable pulsations up to the detection limit of their observations, normally of a few milli-modulation amplitude

(mma). Castanheira et al. (2007) showed that at least a few of these stars are really low-amplitude pulsators (down to 1.5 mma).

At the time of this review, the evidence suggests that the DAV instability strip is pure (see Figure 6). When high-S/N spectra are used to measure the atmospheric parameters and high-S/N timeseries photometry are combined, all stars within the instability strip ($12,270\,\mathrm{K} \geq T_{\mathrm{eff}} \geq 10,850\,\mathrm{K}$) are observed to pulsate. We must ask why we do not find more nonvariables scattered into the instability strip (and variables scattered out, as SDSS J235040.74-005430.8 at 10,370 K $\pm$ 20 K, from two spectra) if the uncertainties really are ~300 K.

Figure 7 shows the DBV instability strip. There are stars not-observed to vary inside the strip, but the temperature and log g determinations are uncertain due to possible contamination by hydrogen.

**7.2. Modeling Convective Efficiency and Strip Boundaries**

Winget, Van Horn & Hansen (1981) show that the positions of the theoretical instability strips depend on the parameterization of the convection zone. Even though Bergeron et al. (1995) show that the atmospheric convection prescription for DAs that fits both the UV spectra and the optical one requires ML2/$\alpha = 0.6$, there is no evidence that such a parameterization works also for the deeper convection zone at the base of the partial ionization zone where pulsation driving occurs. Brassard & Fontaine (1999) calculate the theoretical position of the instability strip with ML2/$\alpha = 0.6$ models but conclude that the position is not consistent with the observed one for temperatures measured from model atmospheres with the same parameterization.

Benvenuto & Althaus (1997) compare the position of the theoretical instability strip as a function of the description of convection (ML2 versus CGM theory) by comparing the thermal timescales. Gautschy & Althaus (2002) calculate the position of the DB instability strip using CGM theory on nonadiabatic models. Both find that the ML2 and CGM methods of modeling convection give equivalent results for the position of the theoretical instability strips.

### 7.3. m-Selection

We saw in section 3.3.1 that rotation, magnetic fields, or any force breaking the spherical symmetry in the star will produce a multiplet with $2\ell+1$ components. If the physical amplitude of the multiplets are the same, the observed relative amplitude will be set by inclination (e.g. Dziembowski 1977, Pesnell 1985 and see Figure 4). We might expect, therefore, to observe $2\ell+1$ components for each mode with spherical harmonic index, $\ell$, and radial overtone number, k. Ideally, one could argue that the physical amplitude of each component of the multiplet should be the same from equipartition, given the essentially identical radial structure of the eigenfunction for each. If this were true, it would be straightforward to measure the inclination of the system from the relative amplitudes: The relative amplitude is a function of the absolute value of m and the inclination angle. The components of each multiplet should have the same relative amplitudes. For reasons not yet understood, this is almost never the case. Why this is not so is one of the outstanding mysteries in the pulsating white dwarf stars. Only in one star, the DOV PG1159-035, do we find multiplets of two values of the spherical harmonic index, $\ell=1$ and $\ell=2$. Even here the relative amplitudes are inconsistent with the equipartition argument unless you average over many multiplets.

In most pulsating white dwarf stars, there is an even greater mystery: we only see singlets, not the expected multiplets. Arguments invoking special inclinations are no longer statistically possible, as there are too many pulsating white dwarf stars now known with only singlet modes. Therefore, we conclude there must be an m-selection mechanism operating in pulsating white dwarf stars; the nature of this mechanism remains to be discovered.

### 7.4. Pulsations in DAs in Cataclysmic Variables

Twelve pulsators were discovered recently in low-mass accretion systems (van Zyl et al. 2004, Nilsson et al. 2006, Mukadam et al. 2007). This implies that mass transfer does not strongly disturb the subsurface partial ionization zone that causes convection and/or pulsation. Accretion raises the external temperature distribution and changes external layers composition, but the underlying structure should be similar to that in single stars. The pulsation properties of models with accretion were investigated by Arras, Townsley & Bildsten (2006); they show the instabilities extend to higher effective temperatures than for pure hydrogen atmosphere single white dwarf stars.

There are also 21 known interacting binary white dwarf stars (IBWDs). These are also called ultracompact binaries, composed of WD+WD or WD+He star. They show pulsations either in the white dwarf or the accretion disk (Solheim et al. 1984, 1998; Provencal et al. 1995; Ramsay et al. 2007; Mukadam et al. 2007). These systems belong to the large class of cataclysmic variables. Asteroseismology of white dwarf stars in cataclysmic variables has not yet been done, but it promises important contributions to the study of accreting white dwarf stars in the future.

## 8. NEW FRONTIERS: MEASURING CONVECTION

In this section and the next, we review the early stages of extensions of asteroseismology beyond the traditionally conceived boundaries of matching observed frequencies with models. We discuss the use of both amplitude and phase information to get more from the data and to exploit increasing observational precision.

Describing energy transport by turbulent convection remains one of the most universally important problems in physics and astrophysics. The landscape of this problem is changing rapidly, with an approach originally suggested by Brickhill (1983) and developed and adapted by Mike Montgomery (2005a,b). Montgomery uses nonsinusoidal light curves of pulsating white dwarf stars to set strong constraints on the nature and properties of turbulent energy transport (convection).

This is a profound development, because convection is ubiquitous in stars and in many other physical situations. Our ignorance of the nature of convection is evidenced by the use in mixing length models of the essentially free parameter "$\alpha$," the ratio of the mixing length to the pressure scale height.

In the outer layers of white dwarfs---as in most stars---the partial ionization of an abundant species is thought to produce the increases in radiative opacity that result in turbulent convection (e.g., Fontaine, Graboske & Van Horn 1977, and references therein). Pulsations provide a way to probe the thickness and depth of the partial ionization zone through the thermal response timescale $\tau_C$ of the convection zone $\tau_C$. The fitting of nearly monoperiodic pulsators is illustrated in **Figure** 8. The lower panels show fits to the light curves of the DBV star PG 1351+489 folded at its main pulsation period. The pulsations perform an important experiment for us: They vary $T_{eff}$ about the star's equilibrium temperature and allow us to determine the equilibrium value of $\tau_C$ and its derivative with respect to $T_{eff}$. This is shown as the solid curves in the upper panels of the figure. The dotted curves are obtained by choosing a value of the mixing length and then calculating $\tau_C$ from equilibrium models as a function of $T_{eff}$. Similarly, Montgomery (2005a,b) shows the results of fits to the folded light curve of the complex pulsator DAV G29-38.

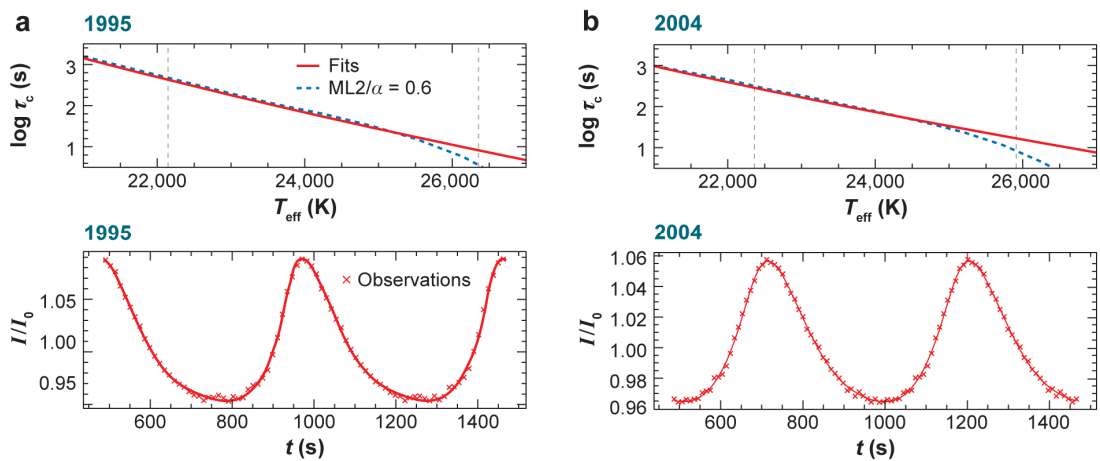

**Figure 8 Nonlinear light curve fits to the nearly monoperiodic pulsator PG1351+489 for two different epochs by Montgomery (2005a). The lower panels show the observations (*red crosses*) and the fits (*solid red line*). Panel a shows a fit using data from a Whole Earth Telescope run in 1995, and panel b is a fit with data from 2004. These light curves have been folded at the period of the main mode, 489 s, to increase the signal-to-noise and simplify the fitting. The upper panels compare the thermal response time scales as a function of $T_{eff}$ which are derived from light curve fits (solid red line) to those which are obtained assuming ML2/α=0.6 convection (dashed blue line).**

Montgomery (2005a,b) also shows that, in addition to the behavior of the convective response timescale, the inclination of the pulsation axis to the line of sight can be measured in this way. Also, it is possible to get mode identifications. Consistently, the application of these techniques to PG 1351 + 489 between 1995 and 2004 showed that $\theta_i$ did not change. This provided the first evidence that the pulsation axis of a white dwarf pulsator is fixed.

Most white dwarf pulsators have more than one mode excited simultaneously. Montgomery (2007a,b) modified his approach to directly use the observed, unfolded light curves. He added the ability to simultaneously fit any number of modes and applied it to the well-studied complex DBV GD 358 as a test case. This star typically has well over 10 modes of different radial overtone number excited to detectable amplitudes. He obtained good fits to its multiperiodic light curve (Montgomery 2007a,b), as shown in Figure 9.

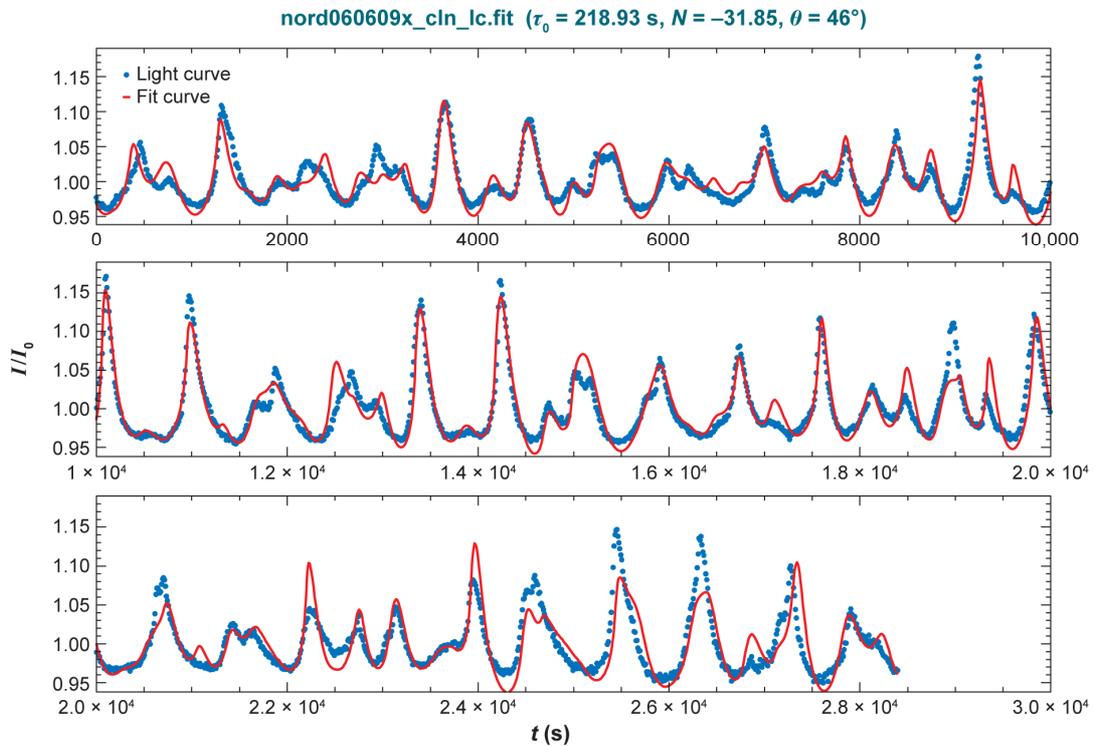

**Figure 9** Fit (*solid red curve*) of the nonlinear pulsation/convection models of Montgomery (2007a, b) to the light curve of GD 358 (blue *points*). The data were taken in May 2006 as part of a Whole Earth Telescope campaign.

These and other data have begun to allow Montgomery and collaborators to characterize the behavior of convection across the DAV and DBV instability strips, but much remains to be done. The fits for large-amplitude pulsators are so good they suggest we understand the variation in the convection zone during a pulsation cycle for these stars, but what about the low-amplitude pulsators and their very different pulse shapes? Also, what about the variation of the depth of the convection through the instability strip? These questions remain to be answered. For GD 358, the observed amplitude of pulsation is so large that the corresponding temperature change completely ionizes the He at the surface, at maximum light. This seems too likely to be an amplitude limiting mechanism (Winget 1998).

## 9. NEW FRONTIERS: PARTICLES AND PLANETS WITH $\dot{P}$

### 9.1. $\dot{P}$ in Multiple Modes: PG 1159 as a Template for the Future

Observations are again moving in front with the measurement of $\dot{P}$ for multiple modes in the same star. The potential science that could come from this is impressive. Costa et al. (2008) extended the mode identifications by Winget et al. (1991) and found 198 pulsation modes for the DOV PG 1159-035, including 29 $\ell=1$ triplets and 46 $\ell=2$ quintuplets, with at least 30 modes with secure identification, and Costa & Kepler (2008) determined the rates of change of the pulsation periods with time, dP/dt, for four modes using the direct measurements of the pulsation periods for the largest amplitude modes, and their detected changes over the 19 years of observations. Using different components of the same triplet, they showed how to measure independently the rate of change of the rotation period, the radius, and the temperature, all weighted by the eigenfunction of the pulsation, which should be extremely similar for different azimuthal components of the same $k$ and $\ell$ mode. This was a demonstration of the power of precision asteroseismology, even though their estimates must be confirmed by observations of the rates of changes of other multiplets. They also measured, for the first time, the second derivatives of the rates of period change for a few modes, showing, for the fast evolving DOVs, that we can achieve such accuracy and challenge the evolutionary models and require the development of new models of stellar rotation and evolution.

**9.1.1. CHANGE IN THE STELLAR ROTATION PERIOD** The observed spacings in frequency between the $m \neq 0$ components and the central ($m=0$) peak, $\nu_m - \nu_0$, of a multiplet are

caused by a combination of the effects of the stellar rotation and of the magnetic field of the star:

$$\delta \nu_m = \nu_m - \nu_0 = \delta \nu_{rot,m} + \delta \nu_{mag,m} \quad (3)$$

To a first-order approximation, the rotation splitting $\delta \nu_{rot,m}$ is proportional to the angular rotation frequency $\Omega_{rot}$ by a factor $m$, while the magnetic splitting $\delta \nu_{mag}$ is proportional to the magnetic strength $B = |\vec{B}|$ by a factor $m^2$,

$$\delta \nu_m \simeq m\, C\, \Omega_{rot} + m^2\, D\, B^2 \quad (4)$$

where $C$ and $D$ are constants. Costa et al. (2008) show that the PG 1159-035 magnetic field strength is very small, $B < 2000$ G, and that the contribution of the average magnetic splitting in the total observed splitting is less than 1% ($\delta \nu_{mag,m} = 0.007 \pm 0.002 \mu Hz$). This allows them to approximate Equation 4 as:

$$\delta \nu_m \simeq m\, C\, \Omega_{rot} \quad (5)$$

The proportionality constant $C$ can be rewritten as

$$C = (1 - C_0 - C_1) \quad (6)$$

where $C_0$ depends on $k$ and $\ell$, $C_0 = C_0(k, \ell)$, and is the uniform rotation coefficient, and $C_1$ is a function of $k$, $\ell$, and $|m|$, $C_1 = C_1(k, \ell, |m|)$, containing the nonuniform rotation effects. This last coefficient depends on the adiabatic pulsation properties, on the equilibrium structure, and on the rotation law. If one assumes uniform rotation, $C_1 = 0$ and $C \simeq 1 - C_0$. In the asymptotic limit with high radial overtones ($k \gg 1$), the uniform rotation coefficient can be approximated (Brickhill 1975) by:

$$C_0 \simeq \frac{1}{\ell\,(\ell+1)} \quad (7)$$

This approximation is good to about $5\%$ for the PG 1159-035 frequency splittings (Costa et al. 2008). For $\ell = 1$ modes, $C_0 \simeq 1/2$ and Equation 5 can be rewritten as

$$\nu_m - \nu_0 \simeq m\frac{1}{2}\Omega_{rot} \qquad (8)$$

Differentiating both sides of the equation with time,

$$\dot{\nu}_m - \dot{\nu}_0 \simeq m\frac{1}{2}\dot{\Omega}_{rot} \qquad (9)$$

or, in terms of periods,

$$\dot{P}_{rot} \simeq 2\left(\frac{\dot{P}_m}{P_m^2} - \frac{\dot{P}_0}{P_0^2}\right)P_{rot}^2 \qquad (10)$$

The accuracy in the determination of $\dot{P}_{rot}$ strongly depends on the accuracies of the $\dot{P}$ s, much more than on the periods. Having two good $\dot{P}$ determinations for a triplet, Costa & Kepler (2008) were able to calculate the rotation period change rate, $\dot{P}_{rot}$, with Equation 10. They used the ($m=0$) 517.1-s and the ($m=+1$) 516.0-s modes and the rotation period, $P_{rot} = 1.3920 \pm 0.0008$ days (Costa et al. 2008), to calculate $\dot{P}_{rot}$:

$$\dot{P}_{rot} \simeq (-2.13 \pm 0.05) \times 10^{-6} \text{ s s}^{-1} \qquad (11)$$

or $\dot{P}_{rot} \simeq -67.2$ s / year. Pre-white dwarf stars such as PG 1159-035 are undergoing a quick atmospheric contraction process. With the contraction, the stellar radius decreases, decreasing the star's angular momentum. To conserve the angular momentum, the angular rotation velocity increases. In other words, its rotation period decreases ($\dot{P}_{rot} < 0$), as suggested by the result above.

**9.1.2. THE CONTRACTION RATE** Costa & Kepler (2008) then calculated the contraction timescale, $R/\dot{R}$, and contraction rate, $\dot{R}$. As shown by Costa, Kepler & Winget (1999), for a star with uniform rotation and negligible mass loss,

$$\frac{\dot{R}}{R} \simeq \frac{1}{2}\frac{\dot{P}_{rot}}{P_{rot}} \qquad (12)$$

Using the result above for $\dot{P}_{rot}$, they obtained

$$\frac{\dot{R}}{R} \simeq (-8.9 \pm 0.2) \times 10^{-12} \text{ s}^{-1} \quad (13)$$

which is equivalent to

$$\frac{\dot{R}}{R} \simeq (-2.8 \pm 0.1) \times 10^{-4} \text{ R}_* \text{ year}^{-1} \quad (14)$$

They used the radius predicted by evolutionary models for PG 1159-035 by Kawaler & Bradley (1994), $R_* = (0.025 \pm 0.005) \ R_\odot$, to estimate the radius change rate $\dot{R}$:

$$\dot{R} = (-2.2 \pm 0.5) \times 10^{-13} \text{ R}_\odot \text{ s}^{-1} \quad (15)$$

or

$$\dot{R} = (-5 \pm 1) \text{ km/yr } \dot{R} = (-5 \pm 1) \text{ km year}^{-1} \quad (16)$$

The contraction rate obtained was

$$\tau_c \equiv -\frac{R}{\dot{R}} = 1.129 \times 10^{11} \text{ s} = 3577 \text{ years.} \quad (17)$$

As pointed out by Costa & Kepler (2008), more realistic calculations of $\dot{R}/R$ must assume differential rotation at least for the outer layer of the star. Models for differentially rotating white dwarf stars calculated by Ostriker & Bodenheimer (1968) suggest that the center rotates faster than the outer layers but never much faster: $\Omega_{\text{surface}}/\Omega_{\text{center}} > 0.2$. In this case, their $\dot{R}/R$ calculated from Equation 12 must be seen as an upper limit for the actual value:

$$\left|\frac{\dot{R}}{R}\right| < 8.9 \times 10^{-12} \text{ s}^{-1} \quad (18)$$

**9.1.3. THE COOLING RATE** Costa & Kepler (2008) could also calculate a first estimate for the PG 1159-035 cooling rate. The changes in period are related with two physical processes in the star, the cooling of the star and the contraction of its atmosphere,

$$\frac{\dot{P}}{P} \simeq -a\frac{\dot{T}}{T} + b\frac{\dot{R}}{R} \quad (19)$$

where $P$ is the pulsation period for the $m=0$ multiplet component, $T$ is the maximum (normally, core) temperature, $R$ is the stellar radius, and $\dot{P}$, $\dot{T}$, and $\dot{R}$ are the respective temporal variation rates (Winget, Hansen & Van Horn 1983). The constants $a$ and $b$ are positive numbers of order unity. Roughly, $a \simeq 1/2$ and $b \simeq 1$ gives

$$\frac{\dot{T}}{T} = 2\left(-\frac{\dot{P}}{P} + \frac{\dot{R}}{R}\right) \quad (20)$$

Using $\dot{P}/P = +2.94\times 10^{-13}\,\mathrm{s\,s^{-1}}$ for the ($m=0$) 517.0 s pulsation mode,

$$\frac{\dot{T}}{T} = (-7.61 \pm 0.50)\times 10^{-11}\,\mathrm{s^{-1}} \quad (21)$$

As $\dot{R}/R = 0.89\times 10^{-11}\,\mathrm{s^{-1}}$, from Equation 13, Equation 19 implies that the PG 1159-035 temporal change in period is controlled by the stellar cooling rather than by the contraction of its atmosphere.

### 9.2. G 117-B15A: Prototype for Mode Identification and $\dot{P}$

G 117-B15A was the first pulsating white dwarf to have its main pulsation mode index identified. The 215-s mode is an $\ell = 1$ mode, as determined by comparing the UV pulsation amplitude measured with the *Hubble Space Telescope* to the optical amplitude (Robinson et al. 1995). Kotak, van Kerkwijk & Clemens (2004), using time-resolved spectra obtained at the Keck Telescope, confirm the $\ell$ measurement for the P = 215-s pulsation but show that the other large-amplitude modes, at 271 s and 304 s, have chromatic amplitude changes that do not fit the theoretical models.

G 117-B15A is proving to be a useful laboratory for particle physics (Isern & García-Berro 2007). Córsico et al. (2001) calculated the limit on the axion mass compatible with the then observed upper limit to the cooling, showing $m_a \cos\beta \leq 4.4\,\mathrm{meV}$. Kepler (2004) demonstrates that, for axion masses of this order, axion cooling would be dominant over neutrino cooling for the lukewarm white dwarf stars. Biesiada & Malec (2002) show that the $2\sigma$ upper limit on the rate of period change published in Kepler et al. (2000b) limits the string mass scale $M_S \geq 14.3 \text{TeV}/c^2$ for six dimensions, from the observed cooling rate and the emission of Kaluza-Klein gravitons, but the limit is negligible for higher dimensions. Benvenuto , García-Berro &Isern (2004) show the observed rates of period change can also be used to constrain the dynamical rate of change $\dot{G}$ of the constant of gravity $\dot{G}$. Bradley (1996, 1998) uses the mode identification and the observed periods of the three largest known pulsation modes to derive a hydrogen layer mass lower limit of $10^{-6}$ $M_*$, and a best estimate of $1.5 \times 10^{-4}$ $M_*$, assuming $k = 2$ for the 215-s mode, and 20:80 C/O core mass. He constrains the core composition from the presence of the 304-s pulsation. Benvenuto et al. (2002) show that the seismological models with time-dependent element diffusion are consistent with the spectroscopic data only if the modes are $\ell = 1$ and k = 2, 3, and 4. They deduce that $M = 0.525 M_\odot$, $\log(M_H/M_*) \geq -3.83$, and $T_{\text{eff}} = 11800$ K, similar to values found by Koester & Allard (2000). Their best model predicts: parallax $\Pi = 15.89$ mas, $\dot{P} = 4.43 \times 10^{-15}$ s s$^{-1}$ for P = 215-s, $\dot{P} = 3.22 \times 10^{-15}$ s s$^{-1}$ for P = 271-s, and $\dot{P} = 5.76 \times 10^{-15}$ s s$^{-1}$ for P = 304-s periodicities, respectively.

The observed rate of period change for the 215s mode of G 117-B15A by Kepler et al. (2005) is therefore consistent with a C or C/O core (see Section 2). The largest uncertainty comes from the models.

It has taken a huge investment of telescope time to achieve the necessary precision, but Kepler et al. (2005) have measured the cooling rate of this 400-million-year-old white dwarf (Wood 1995), excluding the time the star took to reach the white dwarf phase. They have also demonstrated it does not harbor planetary bodies similar to Jupiter in mass up to a distance around 30 AU from the star or smaller planets with light travel time effects on the white dwarf larger than 1 s.

Kepler et al. (2005) claim that the 215-s periodicity in G 117-B15A is the most stable optical clock currently known. Santra et al. (2005) and Hoyt et al. (2005) discuss projects to build optical atomic clocks based on single trapped ions or several laser-cooled neutron atoms of strontium or ytterbium. These are expected to reach an accuracy of $\dot{P} \leq 2\times10^{-17}$ s s$^{-1}$. Considering their periods are $2.5\times10^{-15}$ s and $1.9\times10^{-13}$ s, even though they will be more accurate than G 117-B15A, they will be much less stable, as their timescales for period changes, $P/\dot{P}$, are 125 s and 3 h, compared to 2 Gyr for G 117-B15A. Even the Hulse & Taylor's millisecond pulsar (Hulse & Taylor 1975) has a timescale for period change of only 0.35 Gyr (Damour & Taylor 1991), but the radio millisecond pulsar PSR J1713+0747 (Splaver et al. 2005) has $\dot{P} = 8.1\times10^{-21}$ s s$^{-1}$, and a timescale of 8 Gyr, and PSR B1885+09 = PSR J1857+0943 with $\dot{P} = 1.78363\times10^{-20}$ s s$^{-1}$ has a stability timescale of 9.5 Gyr (Kaspi, Taylor & Ryba 1994), if we neglect the observed starquakes.

**9.3. Axion and Plasmon Neutrino Constraints**

Recent work has focused on combining theoretical models with observations of pulsating white dwarf stars to constrain the mass of axions and the emission rate of plasmon neutrinos (neutrinos that result from the decay of a photon coupled to a plasma). Axions, though hypothetical, are of great interest in astrophysics because they are dark matter candidates. Measuring plasmon neutrino emission rates gives us a unique way to test the theory of weak interactions in the Standard Model of particle physics.

Axions and plasmon neutrinos, if present, would stream freely out of white dwarfs, contributing efficiently to their cooling. We can measure the cooling rate of pulsating white dwarfs by measuring the rate at which the pulsation period of a given mode slows down with time $\dot{P}$. The faster the cooling, the larger $\dot{P}$ is. By comparing the $\dot{P}$ expected from the cooling with the $\dot{P}$ we measure, one can constrain the emission rates of plasmon neutrinos and/or axions and therefore the axion mass.

To date, it has been possible to measure the cooling rate of one white dwarf, the DAV G 117-B15A. Bischoff-Kim, Montgomery & Winget (2007a,b) performed an extensive asteroseismological analysis of G 117-B15A, unprecedented in its scope and precision, and used the results of this analysis and the measured $\dot{P}$ for the

215.2-s mode in G 117-B15A, $(3.57 \pm 0.82) \times 10^{-15}$ s/s $(3.57 \pm 0.82) \times 10^{-15}$ s s$^{-1}$ (Kepler et al. 2005) to place a strong upper limit on the axion mass [Bischoff-Kim, Montgomery & Winget (2008a,b]. **Figure** 10 summarizes the results of this work. With only three observed modes available for analysis, they could constrain the models to regions of parameter space and found two families of models that fit the observed periods to better than 1 s, i.e., the order of uncertainty in the models. Taken together, these models place an upper mass limit on the axion mass of 26.5 meV.

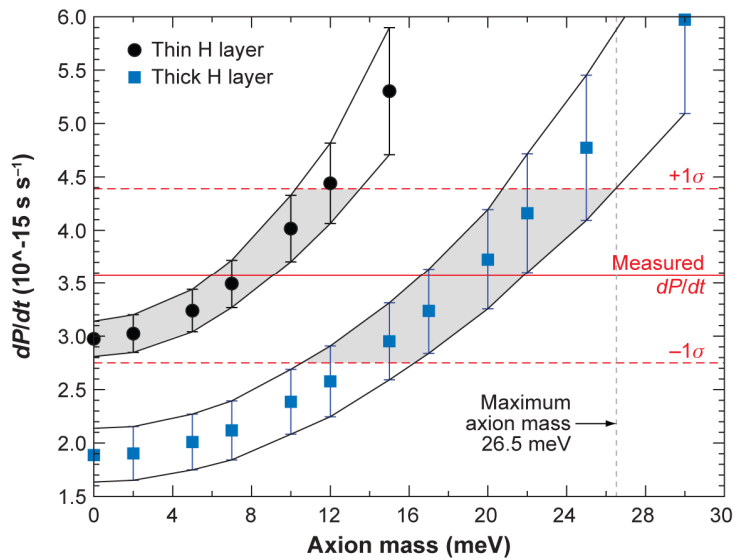

**Figure 10 Axion mass limits derived from Bischoff-Kim, Montgomery and Winget (2008a,b) asteroseismological study of G 117-B15A and the measured $\dot{P}$. The filled symbols with error bars are $\dot{P}$ s calculated for the best-fit models. The error bars were evaluated using Monte-Carlo simulations. They found two families of models that fit the observed periods equally well. Taken together, these models place an upper mass limit on the axion mass of 26.5 meV)**

Based on the same principle, they performed a preliminary asteroseismological analysis of another pulsating white dwarf, EC20058-5234. EC20058 is a hot DBV ($T_{\text{eff}} \sim 28,000$ K, Beauchamp et al. 1999). At this temperature we would expect neutrino emission through plasmon decay to contribute more than half of the cooling (Winget et al. 2004). Sullivan et al. (2007) has collected more than 10 years of data on EC20058, enough to allow a precise measurement of $\dot{P}$ for at least one stable mode. Kim et al. (2006) predicted the limits that we will be able to put on plasma neutrino emission rates once rates of period change become available for EC20058. We present the results of this preliminary analysis in Figure 11.

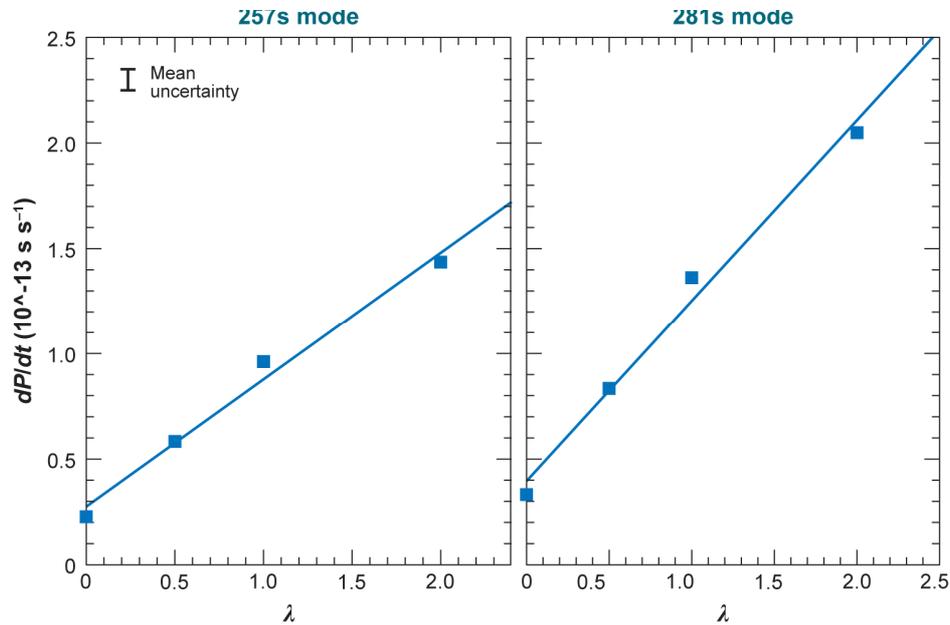

**Figure 11 Limits on plasmon neutrino emission rates based on the preliminary analysis of models for the hot DBV EC20058-5234. $\lambda$ is defined by $\varepsilon_\nu^{measured} = \lambda \varepsilon_\nu^{expected}$. No measurements are available at present. The mean uncertainty indicated in the figure is the expected uncertainty from the measurements, not from the models.**

**9.4. Searching for Planets**

The quantity that we can measure with the greatest precision is time---it can be divided into the smallest intervals. As we have seen in section 9.2, the ability to measure time with an accuracy of fractions of a second has led to the observational demonstration that a subset of pulsating DAV white dwarf stars are the most stable clocks at optical wavelengths (e.g., Kepler et al. 2005)---stable on timescales of gigayears.

If our clock has one or more planetary companions, it will move around the center of mass of the star-planet(s) system. A very stable and precise clock in orbit has proven to be an extremely sensitive and useful scientific instrument in many contexts (e.g., Hulse & Taylor 1975). The orbital motion creates periodic changes in the pulse arrival times of the clock as seen from Earth, corrected to the barycenter of the Solar System. Wolszczan & Frail (1992) detected the first extrasolar planets orbiting a pulsar using this technique.

Even before a full orbit is observed, curvature in the $(O-C)$ diagram gives an early warning of a possible planet through the implied rate of period change, as described in Section 3.4.3.

This technique amounts to light-travel-time astrometry. Because of this, the technique is most sensitive to massive planets at distances comparable to the outer planets in our own Solar System. That is, we can search for planetary systems that are dynamically similar to our own.

Winget et al. (2003) outlined a plan to use this approach for a search for planets around white dwarf stars. The first results from the early stages of this survey are reported by Fergal Mullally and collaborators (Mullally 2007; Mullally, Winget & Kepler 2007). Mullally points out that the limits on planetary companions for two pulsating white dwarf stars, G 117-B15A and R 548, are among the most stringent ever measured. Perhaps the most exciting result to-date from Mullally's survey is the possibility of a Jupiter-mass planet, with an orbital period of about four to five years, orbiting the DAV GD 66 (Mullally et al. 2008).

Interestingly, the Doppler spectroscopic method for finding extrasolar planets has produced the overwhelming majority of all extrasolar planets yet detected. But its sensitivity is very much complemetary to the white dwarf method: It is sensitive to massive planets tht live close to these stars. Mullally's initial survey of only 15 stars may help answer the question, Are planetary systems like our own common or rare?

Summary points
- White dwarf stars comprise up to 98% of the end state of all stars.
- Pulsations are global and sample almost the whole interiors of white dwarf stars.
- White dwarf stars are excellent laboratories for extreme and exotic physics.
- The seismologically determined masses are more accurate than those obtained from binary solutions.

- Asteroseismology is the only tool to measure the surface layer masses, which are determined from the up-to-now not accurately modeled mass loss through stellar evolution.
- Asteroseismology can determine the core composition of white dwarf stars and help to measure the C($\alpha,\gamma$)O reaction rate. The latter cannot be measured in a terrestrial laboratory.
- Asteroseismology can accurately measure the nature and extent of surface partial ionization zones and probe convective energy transport.
- The rates of change of the pulsation periods are measurable and can be used to precisely measure the evolutionary rates of these old stars, to detect planets around them, and to probe for exotic particles that are strong candidates for dark matter.

## FUTURE ISSUES1. PARAMETERS AND PROPERTIES OF THE EQUILIBRIUM STARS

- Are the three instabilities strips pure and the pulsators otherwise normal? Do the pulsations result only from evolution of a star into the region of partial ionization of the dominant surface element or are there other parameters, e.g., mass, surface-layer thickness, and metallicity, that are discriminators? Are all reported nonvariables inside the strips simply low-amplitude pulsators?
- Why is there a low $T_{\text{eff}}$ pulsator in the SDSS sample? Is it just an effect of the low S/N spectra giving uncertain temperatures or is there a physical effect, such as surface layer thickness or metallicity making it an outlier?
- How thin are the surface layers? The evidence points to layers thinner than canonical evolutionary models predict but thicker than the extremely thin H-layers required to mix through convection.
- Are there pulsators in globular clusters and other halo populations? How, if at all, are they different?
- What are the effects of different progenitor metallicities (different populations) on the resulting white dwarf stars? If accreting white dwarf stars are SNIa progenitors when they explode near the Chandrasekhar limit, what are the effects of different metallicities on their peak luminosities? Can we measure the signature of Ne in the core of pulsating white dwarf stars? What are their effects on SNIa luminosities?
- The highest mass DAV pulsators should be crystallized. Are the pulsations coherent or does their inferred restriction to the outermost uncrystallized layers imply short stability timescales?

**2. Mode Properties**
- What are the driving, mode selection, and amplitude limiting mechanism(s)? Are they the same for all strips and throughout each strip? What is the cause of the amplitude changes on timescales from weeks to years, as in GD 358 and G 29-38? For low-mass stars, will p-modes have detectable amplitudes? What is the origin, role, and nature of mode coupling? What is the role of inclination in m-selection, considering we see the amplitude of different m components change with time in a few stars?
- Can we measure the velocity and line profile variations needed for mode identification, necessary for full asteroseismic analysis? Are there other values of the spherical harmonic degree, $\ell$, besides 1 and 2 already observed? Are the modes detected only in the X-ray high of spherical harmonic degree, $\ell$?
- Is driving different for different pulse shapes, or for the DOVs, where the models do not indicate significant convection?

**3. Rates of Period Change**
- Will dark matter give up the secret of its nature? How should we best constrain weakly interacting particles, thereby setting useful constraints on dark matter and electro-weak theory?
- What is the limit, intrinsic to the star, for stability in the most stable pulsators?
- How can we exploit a lattice of very stable clocks through the disk of the Galaxy for physical measurements and technology?

**DISCLOSURE STATEMENT**

The authors are not aware of any biases that might be perceived as affecting the objectivity of this review.

**ACKNOWLEDGMENTS**

D.E.W. and S.O.K. are fellows of Conselho Nacional de Desenvolvimento Científico e Tecnológico-CNPq, Brasil. D.E.W. acknowledges support from the NSF and NASA. The authors thank their mentors: R. Edward Nather, Hugh M. Van Horn, Carl J. Hansen, Edward L. Robinson, and Gilles Fontaine, who are also colleagues and friends. They also thank all their students and former students, also colleagues and friends. They thank Mike H. Montgomery, Agnes Bischoff-Kim, Fergal Mullally, and Barbara Castanheira for helpful comments on early versions of this manuscript, and the carefull editing by John Kormendy, Associate Editor, Roselyn Lowe-Webb, Production Editor, and Douglas Beckner, Senior Illustration Editor.